\pdfoutput=1
\documentclass[sigconf]{acmart}

\renewcommand\footnotetextcopyrightpermission[1]{}
\AtBeginDocument{%
  }
\copyrightyear{2023} 
\acmYear{2023} 
\setcopyright{rightsretained}
\acmConference[ACSAC '23]{Annual Computer Security Applications Conference}{December 4--8, 2023}{Austin, TX, USA}
\acmBooktitle{Annual Computer Security Applications Conference (ACSAC '23), December 4--8, 2023, Austin, TX, USA}
\acmDOI{10.1145/3627106.3627111}
\acmISBN{979-8-4007-0886-2/23/12}

\usepackage{xurl}
\usepackage[absolute,overlay,showboxes]{textpos}

\begin{document}

\title{\textsc{PhishReplicant}: A Language Model-based Approach to Detect Generated Squatting Domain Names}

\settopmatter{authorsperrow=4}
\author{Takashi Koide}
\email{takashi.koide@global.ntt}
\orcid{0009-0008-1942-0335}
\affiliation{%
  \institution{NTT Security (Japan) KK}
  \city{Tokyo}
  \country{Japan}
}
\author{Naoki Fukushi}
\email{naoki.fukushi@global.ntt}
\orcid{0009-0007-8103-8244}
\affiliation{%
  \institution{NTT Security (Japan) KK}
  \city{Tokyo}
  \country{Japan}
}
\author{Hiroki Nakano}
\email{hi.nakano.sec@gmail.com}
\orcid{0009-0009-4470-2139}
\affiliation{%
  \institution{NTT Security (Japan) KK}
  \city{Tokyo}
  \country{Japan}
}
\author{Daiki Chiba}
\email{daiki.chiba@ieee.org}
\orcid{0000-0002-7532-6633}
\affiliation{%
  \institution{NTT Security (Japan) KK}
  \city{Tokyo}
  \country{Japan}
}
\author{}

\renewcommand{\shortauthors}{Koide et al.}

\begin{abstract}

Domain squatting is a technique used by attackers to create domain names for phishing sites.
In recent phishing attempts, we have observed many domain names that use multiple techniques to evade existing methods for domain squatting.
These domain names, which we call \textit{generated squatting domains} (GSDs), are quite different in appearance from legitimate domain names and do not contain brand names, making them difficult to associate with phishing.
In this paper, we propose a system called \textsc{PhishReplicant} that detects GSDs by focusing on the linguistic similarity of domain names.
We analyzed newly registered and observed domain names extracted from certificate transparency logs, passive DNS, and DNS zone files.
We detected 3,498 domain names acquired by attackers in a four-week experiment, of which 2,821 were used for phishing sites within a month of detection. 
We also confirmed that our proposed system outperformed existing systems in both detection accuracy and number of domain names detected.
As an in-depth analysis, we examined 205k GSDs collected over 150 days and found that phishing using GSDs was distributed globally.
However, attackers intensively targeted brands in specific regions and industries.
By analyzing GSDs in real time, we can block phishing sites before or immediately after they appear.

\end{abstract}


\maketitle

\setlength{\TPHorizModule}{\paperwidth}
\setlength{\TPVertModule}{\paperheight}
\TPMargin{8pt}
\begin{textblock}{0.8}(0.1,0.02)
	\noindent
	If you cite this paper, please use the following reference:\\
        Takashi Koide, Naoki Fukushi, Hiroki Nakano, and Daiki Chiba. 2023. PhishReplicant: A Language Model-based Approach to Detect Generated Squatting Domain Names. In \emph{Annual Computer Security Applications Conference (ACSAC ’23)}, December 4–8, 2023, Austin, TX, USA. ACM, New York, NY, USA, 13 pages. https://doi.org/10.1145/3627106.3627111
\end{textblock}

\section{Introduction}

Phishing sites use various techniques to trick victims into believing they are legitimate.
Attackers create fake login pages by copying from the web content of branded websites~\cite{SubramaniMSVP22}.
They also acquire domain names similar to legitimate ones~\cite{AgtenJPN15}.
Users may mistakenly identify these domain names displayed in emails or short message service (SMS) messages as legitimate, click on the links, and unknowingly get redirected to phishing sites~\cite{Liu0LLDS21,TangML0022}.
As a result, users may unknowingly enter sensitive information, such as credit card numbers and login credentials, under the false belief that they are on a legitimate website~\cite{PengXQ0VW19}. 

Previous studies have investigated domain squatting since the 2000s~\cite{WangBWVD06}.
Researchers proposed methods for detecting domain names that differ slightly from legitimate domain names (e.g., typosquatting~\cite{AgtenJPN15,SzurdiLKNC21}, bitsquatting~\cite{NikiforakisAMDPJ13}, and IDN homograph~\cite{Suzuki0YMG19}) and include brand names (e.g., combosquatting~\cite{KintisMLCGPNA17}).
In recent phishing attempts, many domain names are created by using multiple squatting techniques to evade existing detection methods.
For example, these domain names set typosquatting strings of the legitimate domain names to their subdomains and append other words connected by hyphens.
The edit distance between legitimate domain names and most of these domain names is large.
These domain names do not directly contain brand names, making it difficult to associate them with legitimate ones. 
Also, there are strong similarities between these domain names, which appear to be generated by using common techniques.
We refer to these domain names as \textit{generated squatting domains} (GSDs).
Attackers generate a large number of similar domain names by using multiple squatting techniques, such as adding, deleting, or substituting words, letters, or random strings, based on legitimate domain names.
Although they appear to be related by human cognitive characteristics, their similarity is difficult to express on a rule basis, and new patterns emerge frequently. 
Therefore, creating rules to cover all possible patterns requires a significant amount of effort.

In this paper, we propose a system called \textsc{PhishReplicant} that detects domain names registered by attackers. 
Our focus is on identifying GSDs that bear linguistic similarities to \textit{known phishing domain names}. 
To achieve this, we leverage a Transformer-based language model and automatically generate matching rules.
Since \textsc{PhishReplicant} only analyzes domain name strings, we can utilize various types of data without being restricted by data format.
By extracting GSDs from the latest known malicious domain names, \textsc{PhishReplicant} can efficiently detect similar GSDs from the stream of newly registered domain names in real-time. As a result, we can keep up with the latest attacks without the need for manually adding newly targeted domain names or training the model on newly emerged squatting techniques.
By applying \textsc{PhishReplicant} to these data, we can collect GSDs before they are used as phishing sites or early in the emergence of phishing sites.
The source code for \textsc{PhishReplicant} is available at \urlstyle{tt}{\url{https://github.com/tkoide398/PhishReplicant}}.

We conducted a four-week real-time evaluation experiment using Certificate Transparency (CT) logs~\cite{ctlog}, lists of registered domain names, and passive DNS traffic observed on 66 DNS cache servers in 18 countries.
We found 3,498 GSDs registered by attackers, of which 2,821 were used as phishing sites within one month of detection. 
We also conducted experiments comparing \textsc{PhishReplicant} with existing systems, including rule-based and machine learning-based systems for detecting phishing-related domain names.
The results showed that \textsc{PhishReplicant} had the highest detection accuracy and detected the largest number of domain names that did not directly contain exact brand names.

Moreover, we clustered 205k GSDs collected from the results of our proposed system and threat intelligence in 150 days as an in-depth analysis to reveal phishing tactics using GSDs. 
Since most GSDs in the same clusters are used for multiple days (with a median duration of 41 days), identifying GSDs based on the similarity of known phishing domain names can prevent the spread of phishing. 
We also found that phishing using GSDs targets brands in 35 countries and is biased toward certain geographic regions. 
Attackers used GSDs to deploy phishing sites targeting 265 brands, including many financial institutions such as banks and credit card services, to gain a monetary benefit.

In summary, we make the following contributions:
\begin{itemize}
\item We propose a system named \textsc{PhishReplicant} that utilizes language models and automated matching rules to detect GSDs by analyzing the similarity between phishing domain names.
\item We conducted a real-time experiment using \textsc{PhishReplicant} for four weeks and successfully discovered 3,498 GSDs that attackers were likely to have acquired. Among them, 2,821 GSDs were used as phishing sites within one month after detection.
\item We performed a comparative experiment between \textsc{PhishReplicant} and existing systems for detecting phishing-related domain names. The experiment showed that \textsc{PhishReplicant} achieved the highest detection accuracy and identified the most domain names not containing exact brand names.
\item We analyzed 205k GSDs collected over a period of 150 days. The analysis revealed that phishing attacks using GSDs targeted 265 brands in 35 countries. Additionally, we clarified that each cluster of GSDs was observed over multiple days, the distribution of phishing attacks was biased, and attackers frequently targeted financial brands.

\end{itemize}

\section{Generated Squatting Domain}
\label{sec:background}

\begin{figure}[!t]
\centering
\includegraphics[width=\linewidth, bb=0.000000 0.000000 846.000000 399.000000] %
{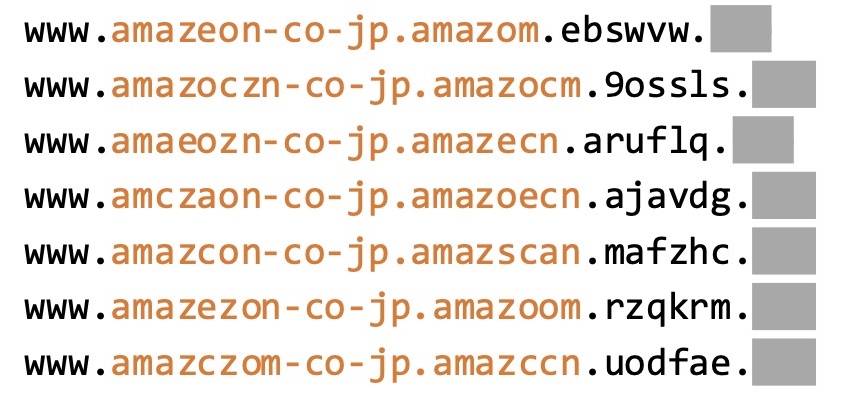}
\caption{Examples of generated squatting domains.}
\label{fig:example_gsd}
\end{figure}

Attackers often obtain domain names that imitate legitimate services by employing domain squatting techniques. 
These domain names are then utilized to create phishing sites or embed links in emails and text messages. 
Previous studies proposed various methods to find domain squatting, e.g., using dictionaries such as ``0'' to ``o'' to create typosquatting~\cite{DamKBS19,SabahNBC22,SzurdiLKNC21,XiaNK0Y21} from legitimate domain names, generating similar domain names by machine learning models~\cite{LoyolaGKWS20,SimpsonMC20}, and detecting combosquatting~\cite{KintisMLCGPNA17,ZengCZT21}, which involves combining a popular brand or trademark with other words or phrases.
We have observed that attackers frequently generate numerous similar domain names, which we refer to as Generated Squatting Domains (GSDs), to circumvent existing countermeasures. Figure~\ref{fig:example_gsd} illustrates examples of GSDs employed for phishing sites resembling \urlstyle{tt}{\url{www.amazon.co[.]jp}}. These GSDs are generated through three primary techniques: combosquatting, typosquatting, and the use of deceptive subdomains~\cite{OestSDAWW18}, which position legitimate domain names within subdomains. 
These GSDs have different second-level domains comprised of six different characters. 
They consist of strings with a maximum edit distance of three from \urlstyle{tt}{\url{amazon}} and strings with a maximum edit distance of three for \urlstyle{tt}{\url{amazon-co-jp}}, a variant of \urlstyle{tt}{\url{amazon.co[.]jp}}. 
In our experimental analysis detailed in Section~\ref{sec:measurement}, we identified 822 domain names similar to these GSDs used for \urlstyle{tt}{\url{amazon.co[.]jp}} phishing sites. While we can recognize the similarities by comparing these GSDs, it is challenging to represent the relationship between GSDs and the original domain names using a straightforward rule-based algorithm, such as the Damerau-Levenshtein distance. For instance, consider a rule designed to identify domain names with an edit distance of one to three between their subdomains and the term ``\urlstyle{tt}{\url{amazon}}.''
When we applied this rule to domain names extracted from Certificate Transparency (CT) logs over a single day, we discovered over 4,500 false positives.
In addition to the examples provided, a variety of patterns exist for generating GSDs, including substituting and adding words or letters, based on multiple domain squatting techniques. Consequently, considerable effort is required to manually deduce how these domain names are generated and formulate rules for matching them.
Similarly, when using machine learning-based methods~\cite{streamingphish,DrichelDBM21} to detect GSDs, numerous false positives may occur, and the detection results may not be comprehensive (we will explain this analysis result in detail in Section~\ref{sec:comparative}).

\section{PhishReplicant}

\begin{figure}[!t]
\centering
\includegraphics[width=\linewidth] %
{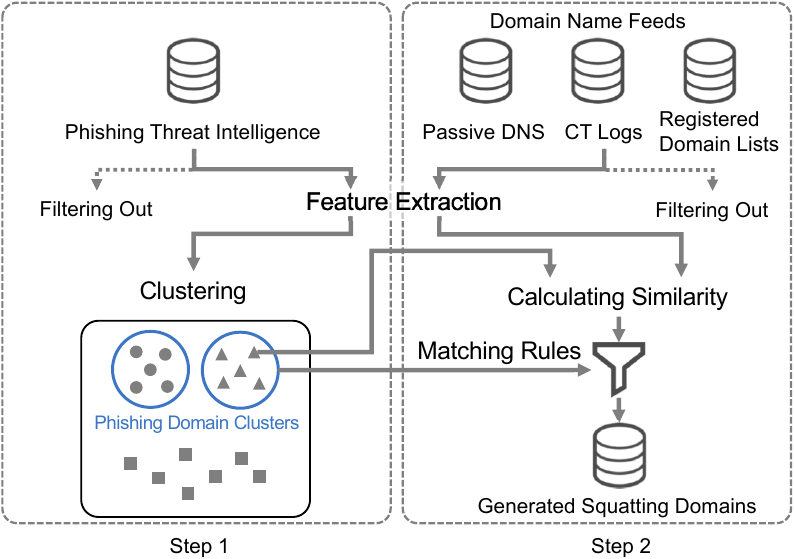}
\caption{An overview of \textsc{PhishReplicant}.}
\label{fig:overview}
\end{figure}

We propose a system called \textsc{PhishReplicant} for detecting GSDs from newly registered or observed domain names as input data.
\textsc{PhishReplicant} analyzes the similarity between domain names of known phishing sites.
Using a Transformer-based language model, the system can capture the linguistic characteristics of domain names and detect GSDs regardless of individual squatting techniques.

An overview of \textsc{PhishReplicant} is shown in Figure~\ref{fig:overview}. 
The system consists of two steps. 
Step 1 takes known domain names used for phishing sites as input and extracts similar domain names through clustering.
At this time, \textsc{PhishReplicant} converts the domain name strings into embedding vectors, which capture the linguistic features of GSDs.
These extracted domain names are subsequently used in Step 2 for comparison purposes. 
This allows the system to extract new GSD patterns that frequently emerge from the latest phishing activity.
In addition, as new brands are targeted by attackers, the system can handle them without manually adding them.
\textsc{PhishReplicant} can even detect GSDs that do not explicitly contain the brand name itself or its variations by extracting patterns from a variety of GSDs without directly comparing them to legitimate domain names.
Step 2 takes newly registered or observed domain names as input. 
The domain names that are similar to those extracted in Step 1 are output as GSDs.
By detecting GSD candidates at the domain name registration and certificate issuance stages, before phishing sites are deployed, we can proactively prevent phishing attacks.

\subsection{Step 1. Extracting Sets of Similar Domain Names}
\label{sec:step1}

We obtain a set of similar domain names from Phishing Threat Intelligence (phishing TI). 
Phishing TI we used in this study comprises three sources: PhishTank~\cite{phishtank}, OpenPhish~\cite{openphish}, and Twitter posts. 
To gather domain names related to phishing from the Twitter stream, we used CrowdCanary~\cite{nakano2023canary}.
CrowdCanary extracts 104 dimensional features from the text and images of Twitter posts searched by keywords such as ``phishing'' and ``email''. Then, CrowdCanary classifies these posts into two classes, phishing-reports and non-reports, using a supervised machine learning model.
We received data on domain names in phishing-reports from the author of CrowdCanary.
To extract domain names used for phishing sites from these domain names, we automatically access the domain names using our crawler as described in Section~\ref{sec:eval2}.
If the accessed web page contains a logo image of a specific brand and its domain name is not that of a legitimate site, our crawler adds the domain name to the Phishing TI.
In the following, we explain ways of excluding unnecessary domain names, extracting features from them, and clustering.

\subsubsection{Filtering Out} 
Phishing TI and newly registered/observed domain names often contain domain names for which the website owner cannot set strings, i.e., domain names that are not used for squatting.
For example, web hosting providers generate large numbers of domain names every day for specific purposes, such as service identification and reverse lookup.
To reduce processing time and false positives, \textsc{PhishReplicant} excludes such domain names in two ways:
\begin{itemize}
\item \textsc{PhishReplicant} excludes domain names that contain generated identifiers in their subdomains (e.g., \urlstyle{tt}{\url{f81d4fae-7dec-11d0-a765-00a0c91e6bf6.example[.]com}}) by checking if one of the subdomains contains a universally unique identifier (UUID) or 32-character hexadecimal number.
\item \textsc{PhishReplicant} excludes domain names whose subdomain names contain IPv4 addresses. It also finds domain names that use hyphens instead of dots to represent IPv4 addresses (e.g., \urlstyle{tt}{\url{ip192-168-100-1.hosting.example[.]com}}).
\end{itemize}
Attackers may create such long squatting domain names that include brand names and UUIDs or IP addresses to evade detection.
However, these lengthy domain names might weaken the effectiveness of domain squatting, as users may become suspicious and avoid interacting with them, reducing the chances of successful attacks.
Since we did not find any such domain names during the period of this study, we excluded them from the scope of this study.
Also, our system excludes domain names that are less than seven characters after removing the top-level domain (TLD) from the fully qualified domain name (FQDN) and those that are all numbers after removing the TLD.

\subsubsection{Extracting Features from Phishing TI}

\textsc{PhishReplicant} utilizes a language model to convert domain name strings into text embedding vectors that represent the hierarchical structure of domain names and combinations of brand names and words.
While simple algorithms such as the Damerau-Levenshtein distance can detect domain names that differ slightly from legitimate ones (e.g., typosquatting), they cannot identify GSDs since GSDs involve substituting, inserting, and deleting words as well as letters.
Even if humans recognize sets of GSDs as similar domain names, simple algorithms cannot express that similarity.

To address this issue, we explore ways to analyze textual similarities between domain names using a language model such as BERT~\cite{DevlinCLT19}.
BERT is a Transformer-based language model that has shown great effectiveness in various natural language processing tasks, such as sentence classification.
When identifying GSDs using BERT, one approach is to add a fully connected layer on top of BERT to perform a classification task, such as binary classification of GSDs and benign domain names, or multi-class classification of targeted brands.
However, this approach has two problems. 
First, analyzing a vast number of domain names and classifying them into the GSD class or brand-specific classes will result in numerous false positives.
Second, new patterns of GSDs emerge frequently, requiring periodic fine-tuning of the BERT model.
Similar issues have been discussed in a previous study~\cite{LinLDNCLSZD21}, which computes the representative vector of the logo image of the phishing site for brand identification and examines the cosine similarity.

Therefore, \textsc{PhishReplicant} uses a language model as a feature extractor rather than a classifier to identify GSDs by calculating the similarity of embedding vectors obtained from domain names. 
Specifically, \textsc{PhishReplicant} uses Sentence-BERT (SBERT)~\cite{ReimersG19}, a modification of the pre-trained BERT network, to represent the structure of domain names in text embeddings.
The original BERT is unsuitable for semantic textual similarity tasks and clustering, which causes a significant computational overhead.
To address this issue, SBERT uses siamese and triplet network structures to derive semantically meaningful text embeddings.
We fine-tune SBERT on a dataset containing GSDs by using the triplet loss function to comprehend domain name similarities.
Please refer to Section~\ref{sec:training} for the specific training process of SBERT.

\subsubsection{Clustering}
\label{sec:clustering}

\textsc{PhishReplicant} clusters domain names using DBSCAN to extract sets of similar domain names. 
DBSCAN is a clustering algorithm that determines clusters by specifying the maximum distance of the neighborhood and the minimum number of data points within the clusters. 
DBSCAN outputs clusters on the basis of neighborhood density so that similar domain names can be grouped into the same cluster. 
Also, unlike clustering algorithms such as k-means, DBSCAN does not depend on the shape of the clusters or require specifying the number of clusters. 
Domain names not labeled into any clusters can be excluded as noise. 
We use cosine similarity as distance metrics of DBSCAN. 
We set the minimum number of data points ($MinPts$) to three and the maximum distance ($eps$) to 0.04 on the basis of the experiment in Section~\ref{sec:eval1}.

\subsubsection{Generating Matching Rules}

GSDs may share common characteristics, including the use of identical top-level domains (TLDs), effective second-level domains (e2LDs), such as \urlstyle{tt}{\url{example.co[.]uk}}, or the same number of characters.
To minimize false positives, \textsc{PhishReplicant} leverages these characteristics to generate matching rules for each cluster.
Specifically, \textsc{PhishReplicant} creates three types of rules by examining whether all domain names within the cluster share the same TLD, e2LD, or number of characters. If any of these conditions are met, the system outputs a matching rule that is common to all domain names within the cluster. For instance, suppose there is a cluster containing domain names \texttt{example000.test}, \texttt{example001.test}, and \texttt{example002.test}. In this case, the output rule would be
\texttt{\{"tld":".test","num":15\}}.

\subsection{Step 2. Detecting GSDs from Newly Registered and Observed Domain Names}
\label{sec:step2}

In Step 2, \textsc{PhishReplicant} receives newly registered and observed domain names and detects GSDs similar to known phishing domain names extracted in Step 1.

\subsubsection{Extracting Features from New Domain Names}

\textsc{PhishReplicant} receives new domain names collected from data such as CT logs, passive DNS traffic, and lists of registered domain names.
For CT logs, we only consider domain names that have not been previously observed because CT logs include not only new domain registrations but also certificate renewals. Similarly, we also extract only newly observed domain names from passive DNS traffic.
Since \textsc{PhishReplicant} uses only strings of domain names for similarity calculation, we can input various data without depending on the data format.
Furthermore, our system can detect GSDs immediately after domain name registration and certificate issuance.
\textsc{PhishReplicant} also filters out domain names whose subdomains cannot be set by users and extracts features in the same way as in Step 1.

\subsubsection{Calculating Similarity}

\textsc{PhishReplicant} calculates the similarity between new domain names and domain names clustered in Step 1.
we consider two sets of embedding vectors denoted by $\mathcal{N}$ and $\mathcal{P}$, respectively. Set $\mathcal{N}$ comprises of $\boldsymbol{u}_1, \boldsymbol{u}_2, ..., \boldsymbol{u}_m$, which represent embeddings for new domain names. Set $\mathcal{P}$ comprises of $\boldsymbol{v}_1, \boldsymbol{v}_2, ..., \boldsymbol{v}_n$, which correspond to embeddings for clustered domain names from Phishing TI.
The cosine similarity between any two vectors $\boldsymbol{u}_i$ and $\boldsymbol{v}_j$ is defined as follows:
\begin{equation}
\cos(\boldsymbol{u}_i,\boldsymbol{v}_j) = \frac{\boldsymbol{u}_i \cdot \boldsymbol{v}_j}{\lVert \boldsymbol{u}_i \rVert \lVert \boldsymbol{v}_j \rVert}.
\end{equation}
For each vector $\boldsymbol{u}_i$ in $\mathcal{N}$, we calculate its cosine similarity with all vectors in $\mathcal{P}$. 
After calculating the cosine similarity between each vector $\boldsymbol{u}_i$ in $\mathcal{N}$ and all vectors in $\mathcal{P}$, we examine the number of vectors in $\mathcal{P}$ that exceed a threshold value $t$. This threshold value is defined as 1 minus the cosine distance value ($eps$) used in Section~\ref{sec:clustering}, which is set to 0.96. If the number of vectors in $\mathcal{P}$ that exceed the threshold value $t$ is equal to or greater than $k$, we output the corresponding domain name as a candidate of GSDs.
We set the $MinPts$ to 3 for clustering, so we set $k$ to a smaller value of 2.

If this calculation is performed on a brute-force basis, the time complexity of this calculation represents $O(mn)$, where $m$ is more than 1M for new domain names, and $n$ is more than 50k for known phishing domain names for one day of our experiment.
To reduce this enormous amount of calculation, we use Faiss~\cite{faiss}, a library for efficient similarity search of dense vectors.
Faiss efficiently finds k-nearest neighbors of a query vector in large, high-dimensional vector databases through approximate search.
After \textsc{PhishReplicant} finds domain names, the system applies the matching rules to each candidate if the most similar domain name has corresponding rules.
Our system excludes the domain names that do not match the rules and outputs the remaining ones as GSDs.

\subsection{Training}
\label{sec:training}

We explain how to train the SBERT model used in Sections~\ref{sec:step1} and~\ref{sec:step2} to represent the similarity of domain names.
In this paper, we use SentenceTransformer~\cite{sbert} framework and fine-tune a pre-trained model (all-mpnet-base-v2).
The advantage of using SentenceTransformer is that it provides simple ways to fine-tune pre-trained models and compute dense vector representations for sentences.
The pre-trained transformer model uses a tokenizer to split sentences into vocabulary tokens and assigns IDs to each of them. 
The model that we fine-tune adopts a method called WordPiece tokenizer, which splits some words into sub-words.
For example, the tokenizer splits \urlstyle{tt}{\url{www.mastercard[.]com}}
into \urlstyle{tt}{\url{www|.|master|###card|.|com}} (\urlstyle{tt}{\url{###}} represents the suffix following words).
Thus, some brand names may not be represented. 
Therefore, we added common TLDs and brand names often abused for phishing sites to the tokenizer's vocabulary.
We chose 1,292 TLDs that appeared in the two months of phishing TI.
To collect brand names, we used the 277 brands that are registered in PhishPedia~\cite{LinLDNCLSZD21}, and we also extracted 105 brands that were recently used in phishing attacks from the OpenPhish and PhishTank feeds. In total, we added 382 brands to the tokens.

The purpose of this training is to create a text embedding model that maps similar squatting domain names into a close vector space.
Domain names can be shorter than the sentences that language models usually handle and have their own structure delimited by dots.
By learning the structure of domain names as train data, the model can accurately identify the similarity of GSDs.
We apply Triplet Loss~\cite{ReimersG19} to fine-tune the SBERT model.
Triplet Loss is a loss function that takes three sentences (anchor, positive, negative) as input and outputs corresponding embedding vectors. 
The objective of this loss function is to minimize the distance (inverse of cosine similarity) between the anchor and positive embedding vectors and to maximize the distance between the anchor and negative embedding vectors. 
In other words, the model learns that the anchor and positive domain names are semantically close, while the anchor and negative domain names are semantically distant.
We create a structure called a triplet network that processes three sentences (triplets) with the same parameters. 
The output of the network is passed to the Triplet Loss function to calculate the loss value. 
We update the parameters to minimize this loss value.
To prepare a triplet dataset, we extracted 42,311 GSDs from two months of phishing TI (the way of selecting GSDs is described in Section~\ref{sec:eval1}) and created 100k pairs of anchors and positives from similar GSDs.
We randomly sampled 100k benign domain names as negatives from CT logs and created 100k triplets.
We used this fine-tuned model, which was trained on the triplet data, in the following experiments.

\subsection{Deploying}

We explain how to deploy \textsc{PhishReplicant} to detect the latest daily GSDs.
Since the domain names in phishing TI are regularly updated, we should run Step 1 with updated data periodically.
In this paper, we collected domain names from phishing TI up to two months ago and ran Step 1 once a day.
Step 2 should be executed as soon as new domain names are collected.
We scheduled a job to detect GSDs once a day, as in Step 1, for our experiment.
If we monitor the stream from CT logs and passive DNS in real time, we could detect GSDs immediately after domain registration and certificate issuance.

\section{Evaluation}

In this section, we describe the evaluation experiments conducted on \textsc{PhishReplicant}. First, we evaluated the clustering results of Step 1 using a dataset. Then, we validated the output of the entire system during a certain period.
Additionally, we conducted an experiment comparing \textsc{PhishReplicant} with existing tools for detecting domain names related to phishing.

\subsection{Evaluation of Clustering}
\label{sec:eval1}

We performed an evaluation experiment to confirm that we can correctly find phishing domain names from phishing TI by clustering in Step 1.

\subsubsection{Creating Dataset}

Since no existing data includes only GSDs and no existing methods identify GSDs, we need to create a dataset for evaluation. 
We labeled 34,095 domain names in the phishing TI for 20 days in October 2022 as positive (GSD) or negative. 
As mentioned in Section~\ref{sec:background}, GSDs have various patterns and are difficult to identify by specific rules. 
Therefore, using the guidelines below, we manually extracted GSDs on the basis of the characteristics, where we can notice similarities between GSDs if they are listed.

We create sets of at least three similar domain names that satisfy all of the following four conditions.

\begin{enumerate}
  \item Domain names contain the common strings: the entire or part of a brand name, or its typosquatting.
  \item The difference in length of the domain name excluding TLD or e2LD is less than three.
  \item If there are more than two subdomains, the number of subdomains is the same.
  \item The number and position of dots and hyphens (if they exist) are the same.
\end{enumerate}

There are some GSDs that we cannot determine on the basis of the above conditions. 
We further consider the following information to add similar domain names to the sets.
\begin{itemize}
  \item URL paths.
  \item The date of domain registration and certificate issuance.
  \item The scan date of VirusTotal~\cite{virustotal} and URLScan~\cite{urlscan}.
  \item Name servers.
  \item Web content of the phishing sites when accessing the domain names.
  \item IP addresses. We exclude some hosting services' IP addresses because they may be shared among multiple users.
\end{itemize}

As a result, we labeled 8,411 (24.7\%) domain names belonging to 1,011 sets for the positive data and 25,684 domain names for the negative data. 
We confirmed that the dataset had been correctly labeled with a review by security experts. Examples of the dataset are shown in the Appendix~\ref{sec:appendix}.

\subsubsection{Clustering}

\begin{table}[!t]
\centering
\caption{Clustering results for each EPS.}
\begin{tabular}{lrrr}
\toprule
$eps$  & Precision & Recall & Accuracy \\
\midrule
0.01 & 100.0\%   & 28.2\% & 81.0\%   \\
0.02 & 99.8\%    & 52.2\% & 87.1\%   \\
0.03 & 99.2\%    & 70.5\% & 91.9\%   \\
0.04 & 96.0\%    & 83.9\% & 94.6\%   \\
0.05 & 89.8\%    & 90.2\% & 94.4\%   \\
0.06 & 82.3\%    & 95.0\% & 92.9\%   \\
0.07 & 75.3\%    & 97.4\% & 90.3\%   \\
\bottomrule
\end{tabular}
\label{tab_cluster}
\end{table}

We calculated embedding vectors from each domain name in the dataset by using the SBERT model trained in Section~\ref{sec:training} and clustered them by DBSCAN.
We performed clustering seven times, changing $eps$, the threshold for the distance between vectors, by 0.01, from 0.01 to 0.07.
Domain names belonging to one of the clusters are identified as positive, and domain names not belonging to any cluster are identified as negative.
Table~\ref{tab_cluster} shows results for each $eps$ for the three indicators: precision, recall, and accuracy.
The higher the $eps$, the more domain names belong to clusters, i.e., reducing false negatives.
However, the fraction of true positives among the clustered instances (precision) is reduced.
In the following experiments, we set 0.04, the $eps$ with the highest accuracy (94.6\%), to perform the clustering in Step 1.
Also, we use this value for the threshold of the cosine distance in Step 2 to detect GSDs.
As true positives, we were able to identify similar domain names where words were substituted, and some letters were changed or added.
For example, we found phishing domain names targeting Apple (\urlstyle{tt}{\url{apple-event-portal-support-online[.]com}}) and targeting Yahoo (\urlstyle{tt}{\url{yahmailllllll.godaddysites[.]com}} and \urlstyle{tt}{\url{yahmailll0.godaddysites[.]com}}).
We also identified domain names with different e2LDs but similar subdomains.
On the other hand, domain names that consist mostly of numbers and meaningless strings were identified as false positives.
Examples are \urlstyle{tt}{\url{3dq3e8b20gln5j27tro7bk7d24155gk8.web[.]app}} and \urlstyle{tt}{\url{hmzl0b2af7b7eed3176e826c7a.web[.]app}}.
This is because the SBERT model splits long numbers into two- or three-character tokens; thus domain names containing those tokens are incorrectly clustered.
However, we correctly identified domain names that consisted of a brand name or its squatting followed by a hyphen, and random numbers and strings.
Examples are phishing domain names targeting a governmental organization in Germany (\urlstyle{tt}{\url{de-agb-session-q4ki42v[.]xyz}} and \urlstyle{tt}{\url{de-agb-session-q3j4na[.]xyz}}) and Facebook (\urlstyle{tt}{\url{business-meta-team-123908233.web[.]app}} and \urlstyle{tt}{\url{meta-business-form-1298067198.web[.]app}}).

\subsection{Evaluation of Real-time GSD Detection}
\label{sec:eval2}

We used \textsc{PhishReplicant} to detect GSDs from the input of newly registered or observed domain names and validated that the GSDs were used for phishing sites.

\subsubsection{Experimental Setup}

We explain how to collect domain names for input and how to determine if a detected GSD was later used as a phishing site.
We collected domain names from entire CT logs and Zonefiles~\cite{zonefiles}, a list of newly registered domain names for four weeks (28 days) starting in November 2022.
Note that we can only extract domain names of phishing sites using server certificates from CT logs.
However, HTTPS phishing attacks have been reported to account for most (85.1\%) phishing campaigns in recent years~\cite{KimCKDSAD21}.
In addition, we confirmed that 48.7\% of entries in phishing TI for a specific three-month period were HTTPS URLs.

We also observe passive DNS traffic as a data source of newly appeared domain names. 
The passive DNS traffic is gathered from 66 DNS cache servers in 18 countries on a global Tier 1 network. 
We extract domain names that have yet to appear from this traffic, where 13M domain names are newly resolved daily. 
Phishing sites may have already used domain names observed in passive DNS for phishing sites. 
Unlike a previous method~\cite{SabahNBC22}, which analyzes the lifetime and traffic of each domain name, our proposed system only analyzes strings, allowing us to take action in the early stages of emerging phishing sites. 

To confirm whether the detected GSD was actually used as a phishing site, we conducted the following five types of verification.

\noindent\textbf{URL Inspection Service:} We use APIs of VirusTotal, URLScan, and Google Safe Browsing to verify that the detected GSD was reported as a phishing attempt. This confirmation process is performed for a maximum of one month.

\noindent\textbf{Phishing TI:} After detecting GSD, we verify whether it is listed on phishing TI for up to one month.

\noindent\textbf{Web Crawling:} We made a web crawler to identify phishing sites on the basis of web content. 
The crawler repeatedly accesses GSDs daily for up to one month immediately after detection. 
We implemented the crawler using NodeJS and Google Chrome as a browser. 
Although Selenium is the most well-known browser automation tool, it could be detected by anti-bot systems~\cite{AzadSLN20}. 
Thus, we used ChromeDevToolsProtocol~\cite{cdp} to automate the browser. 
Some phishing sites do not allow access to web content due to a cloaking technique~\cite{OestSDAWT19,crawlphish} if not accessed with the proper browser environment. 
For example, we observed that some phishing sites delivered via SMS check the browser's UserAgent is mobile (iOS or Android) using JavaScript and either respond with phishing sites or redirects to other URLs (e.g., \urlstyle{tt}{\url{https://www.gooogle[.]com/}}). 
We set UserAgent for Google Chrome on iOS and the viewport size as 390x844 pixels. 
We used Phishpedia~\cite{LinLDNCLSZD21} to identify a logo image in the screenshot to determine whether web content is related to phishing. We trained a Phishpedia model with logo images of 382 brands (the same as Section~\ref{sec:training}). 
Also, we visually check screenshots not identified by this model and confirm they display login forms or statements asking for money, such as requests for credit card numbers. 

\noindent\textbf{Passive DNS:} Phishing sites that target users with specific attributes (e.g., region, device) or are accessible for only a short time may be missed in the above way. 
Therefore, we use passive DNS and confirm IP addresses registered as A records of phishing TI's domain names for the past two months. 
We queried Farsight DNSDB~\cite{dnsdb} and our passive DNS to retrieve IP addresses. 
If a GSD uses an IP address for phishing, we determine that the GSD has been used as a phishing site. 
We exclude some web hosting services (e.g., \urlstyle{tt}{\url{firebaseapp[.]com}}, \urlstyle{tt}{\url{square[.]site}}, and \urlstyle{tt}{\url{godaddysites[.]com}}) because their IP addresses are shared with other users. 

\noindent\textbf{Manual Validation:} We manually check that GSDs are similar to known phishing domain names in the same way as Section~\ref{sec:eval1}. 
GSDs that the security vendors missed may have been running as phishing sites. 
Also, some GSDs not used as phishing sites may have been registered by attackers as part of phishing campaigns. 
To complement those domain names, we performed this manual check.

\subsubsection{Results}

We analyzed newly registered and observed domain names for four weeks using our proposed system and detected 3,784 domain names.
We confirmed through a manual check that 3,498 (92.4\%) domain names were true positives, indicating they were related to known phishing domain names and were most likely obtained by attackers. Therefore, there were 286 false positives, accounting for 7.6\% of the total.
There are 2,821 (74.6\%) domain names that were used as phishing sites after we detected them, which is the total number of those identified through validations, excluding manual checks.
A total of 1,221 domain names, including 934 (VirusTotal), 433 (URLScan), and 564 (Google Safe Browsing) domain names, were actually used for phishing.
In addition, we identified 430 domain names with Phishing TI and 757 domain names by web crawling.
When we crawled a set of similar domain names (resolved to a small number of common IP addresses) from a single source IP address and confirmed a few phishing sites, followed by 404 responses from all the remaining domain names. 
Thus, some phishing sites limited the number of accesses from the same IP address.
Of domain names that were successfully accessed and were not confirmed to have phishing content, websites of 631 domain names exposed the cgi-bin directory~\cite{SabahNBC22}.
Although attackers may have once used these websites for phishing, they already abandoned them after they removed the phishing content.
As a result of analyzing IP address sharing using passive DNS, we confirmed that 2,106 domain names used the same IP addresses as those used by known phishing domain names.
Unless the IP addresses used for phishing sites were coincidentally assigned, attackers retained control of the IP addresses and used them for other GSDs.

\subsection{Comparative Evaluation with Baseline Systems}
\label{sec:comparative}

In the above analysis, we conducted evaluations to determine how many of GSDs detected using \textsc{PhishReplicant} were actual phishing sites, including those not contained in existing phishing TI. Here, we compared the performance of \textsc{PhishReplicant} with baseline systems for detecting domain names related to phishing. We conducted this comparison by running them on the same types of input feeds corresponding to the period from March 1 to March 31, 2023, and then comparing their detection results with phishing TI. We used four existing systems as baselines in our study: dnstwist~\cite{dnstwist}, Phishing Catcher~\cite{phishingcatcher}, StreamingPhish~\cite{streamingphish}, and Ctl-pipeline~\cite{DrichelDBM21}.

\subsubsection{Baseline Systems}

\begin{table*}[!t]
\centering
\caption{Comparison of detection results among \textsc{PhishReplicant} and baseline systems.}
\label{tab_compare}
\scalebox{0.95}{
\begin{tabular}{lrrrrr}
\toprule
System           & \# of detected domains & \begin{tabular}{r}\# of matched domains \\ with phishing TI or GSB \end{tabular} & \begin{tabular}{r} Ratio of matched domains \\ with phishing TI or GSB \end{tabular} &\begin{tabular}{r} \# of matched domains \textbf{NOT} \\  containing exact brand names \end{tabular} \\
\midrule
dnstwist~\cite{dnstwist}         & 352,294                & 3,645                                & 1.0\%                    & 480                                              \\
Phishing Catcher~\cite{phishingcatcher} & 98,326                 & 705                                  & 0.7\%                    & 226                                              \\
StreamingPhish~\cite{streamingphish}   & 196,677                & 3,770                                & 1.9\%                    & 1,237                                            \\
Ctl-pipeline~\cite{DrichelDBM21}     & 50,441                 & 201                                  & 0.4\%                    & 183                                              \\
\textsc{PhishReplicant}   & 7,358                  & 1,923                                & \textbf{26.1\%}          & \textbf{1,800}         \\                      
\bottomrule
\end{tabular}
}
\end{table*}

We describe baseline systems we used to evaluate the performance of \textsc{PhishReplicant}.

\noindent\textbf{dnstwist~\cite{dnstwist}} is a system that generates candidates of squatting domain names based on a given domain name, using their fuzzing algorithms and dictionaries. It supports multiple types of squatting, such as typosquatting, bitsquatting, and homoglyph. Although Dnstwist has a online phishing inspection function for web pages, we used it as a generator of squatting domain names in this study. Seed domain names used as input were the domain names of 382 brands described in Section~\ref{sec:training}. 

\noindent\textbf{Phishing Catcher~\cite{phishingcatcher}} is a system that extracts domain names related to phishing by analyzing common names and SAN fields in server certificates. It calculates a score based on rule-based methods, such as keyword matching and Levenshtein distance comparison with legitimate domain names, and identifies the domain names as malicious. 

\noindent\textbf{StreamingPhish~\cite{streamingphish}} is an machine learning (ML) based system that identifies phishing domain names by inputting domain names, extracting features from various fields of certificates, and domain name strings, and using a classifier (Logistic Regression). We used the attached dataset to train the classifier. 

\noindent\textbf{Ctl-pipeline~\cite{DrichelDBM21}} is an ML-based system that identifies phishing domain names from server certificates by analyzing various certificate fields, domain name strings, and keywords contained in domain names.
It extracts 126 types of features for its classifier. To train the Ctl-pipeline classifier with the latest data, we created a new dataset from CT logs and collected certificates from November 2022 to February 2023. 
We extracted certificates whose domain names were listed on OpenPhish or PhishTank as phishing-related certificates.
A total of 27,153 certificates were labeled as phishing.
We sampled the same number of certificates from the CT logs for benign label data. 
We used the ExtraTreesClassifier as the classifier and trained on the dataset. 
After training the classifier, we obtained a model capable of detecting certificates containing phishing domain names with an Accuracy Score of 0.931 and an ROC AUC Score of 0.981. 
We set the threshold of the probability score for phishing labels to the default setting of 0.925. If the score exceeded this threshold, the domain name of the input certificate was determined to be phishing.

\subsubsection{Results}

Table~\ref{tab_compare} displays a comparison of detection results among \textsc{PhishReplicant} and the baseline systems. 
\textsc{PhishReplicant}, with 1,923 out of 7,358 domain names matching the Phishing TI or Google Safe Browsing, achieved the highest percentage of matching domain names at 26.1\%. 
This is a significantly higher accuracy level than the 0.4 to 1.9\% of results achieved by the baseline systems. Note that the difference between 7,358 and 1,923 does not necessarily mean that all were false positives. In fact, when further investigating the detection results of \textsc{PhishReplicant} using VirusTotal and URLScan, we found that 2,823 (38.4\%) domain names matched in total. Among the domain names detected by \textsc{PhishReplicant}, 1,682 domain names were not detected by any other systems. This accounted for 87.5\% of the domain names detected by \textsc{PhishReplicant}, demonstrating that the proposed system was able to detect many domain names that could not be detected by other baseline systems.

Although the number of matched domain names by the proposed system was lower than dnstwist and StreamingPhish, the results of these baseline systems include a large number of domain names that directly contain brand names without alteration, such as \urlstyle{tt}{\url{www.apple.ifindmy-id[.]com} and \urlstyle{tt}{\url{paypal.em-inff[.]xyz}.
Specifically, dnstwist's output results showed that out of the domain names that matched the phishing TI or GSB, 3,165 (86.8\%) contained exact brand names, and StreamingPhish output 2,533 (67.2\%) matched domain names that contained exact brand names.
These domain names could be easily found by searching for brand names. When examining the number of matched domains that did not include exact brand names, \textsc{PhishReplicant} had the highest count at 1,800. 

Rule-based systems, such as dnstwist and Phishing Catcher, require the input of legitimate domain names that serve as the basis for detecting squatting domain names. Therefore, when a new brand becomes a target of phishing attacks, these systems require manual updates to include the new legitimate domain names. Additionally, ML-based baseline systems, such as StreamingPhish and Ctl-pipeline, need to update and modify their classifiers to align with the latest phishing trends. In contrast, \textsc{PhishReplicant} automatically extracts domain names as GSDs from the latest phishing TI and detects similar domain names from new domain names based on them. This eliminates the need for frequent model updates required by baseline systems.

\section{In-depth Analysis of GSDs}
\label{sec:measurement}

We analyzed the characteristics of sets of similar GSDs in detail, revealing the tactics and the infrastructure used by attackers.
We clustered domain names detected by the proposed system and domain names extracted phishing TI for 150 days from August 2022 in the same way as Section~\ref{sec:clustering}.
We extracted 205,158 domain names (including 3,784 detected GSDs in Section~\ref{sec:eval2}) labeled into 2,842 clusters.
Each cluster consists of an average of 72.2 domain names.
The cluster with the most domain names comprised 79,547 domain names.
There are four clusters containing more than 10k domain names.

\subsection{Duration of GSDs}
\label{sec:duration}

\begin{figure}[!t]
\centering
\includegraphics[width=\linewidth]
{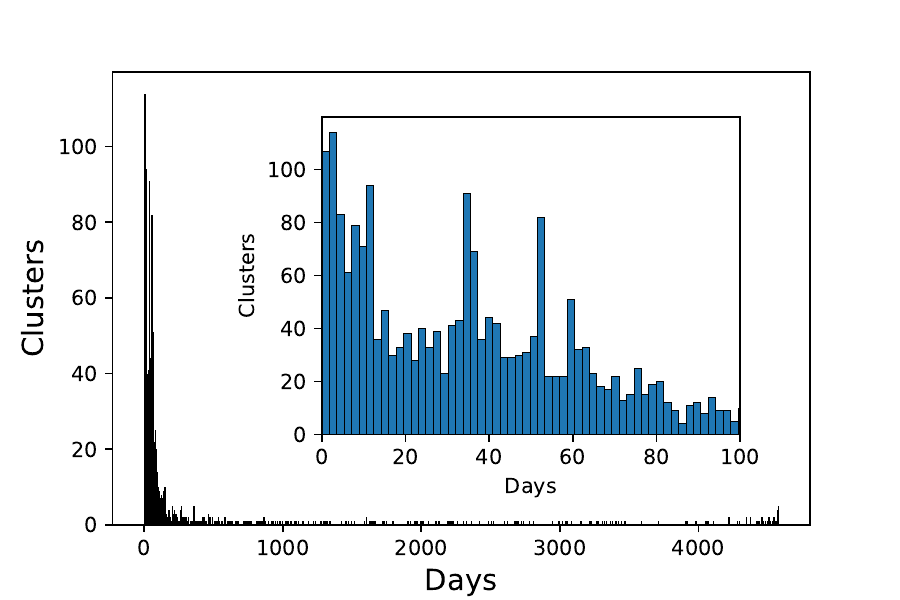}
\caption{Comparison of time differences between the earliest and latest first-seen dates of domain names for each cluster. The inner bars highlight data within the first 100 days, while the outer bars display the overall data.}
\label{fig:duration}
\end{figure}

We analyzed how long GSDs labeled to the same cluster were in use. 
We used the two types of passive DNS described in Section~\ref{sec:eval2} to confirm the dates when domain names were first seen and identify the time difference between them in each cluster, as shown in Figure~\ref{fig:duration}. 
The difference between each cluster's earliest and latest first-seen dates was 262 on average and a median of 41 days. 
If all similar domain names appeared in a short period (e.g., a few hours), our proposed system may not work effectively. 
However, 2,779 (97.8\%) clusters of GSDs are used for more than one day, indicating that we can almost completely prevent access to phishing sites by detecting GSDs on the basis of the similarity of the first ones.

We found that some domain names have been used intermittently for phishing sites over several years.
While GSDs created using multiple squatting techniques are often out of use within a few dozen days, typosquatting domains with close edit distance to legitimate domain names are used for a long time.
For example, \urlstyle{tt}{\url{steamcomnulty[.]com}} has been used frequently to host phishing websites over the past decade and is considered a drop catch, an act of reacquisition of an expired domain name.

\paragraph{Case Study: Cluster A}
We observed a cluster consisting of 4,639 domain names, each of which first appeared over eight days. Examples are \urlstyle{tt}{\url{www.macesaoeod.of6jh4[.]icu}}, \urlstyle{tt}{\url{www.maceseoeod.r7302f[.]icu}}, and \urlstyle{tt}{\url{www.macesarrod.mfxzq4[.]icu}}.
These domain names appear to be squatting of MasterCard; however, they were used for phishing sites of three different credit card brands, which shared a single IP address of ASN-QUADRANET-GLOBAL (AS8100).

\paragraph{Case Study: Cluster B}
We found another cluster consisting of 593 domain names, each of which first appeared over three days. Examples are \urlstyle{tt}{\url{www.vianiocercenure.visoreoecssvxarercmsvi.baflrzt[.]rest}}, \urlstyle{tt}{\url{www.vianioceorcenure.visoreoecssiercmsvi.xppqjxo[.]rest}}, and \urlstyle{tt}{\url{www.vianiocenure.visoresiercmsvi.pefczzw[.]rest}}.
These domain names have different e2LDs as in the clusters above and six patterns in each subdomain.
They were used for phishing sites for two credit card brands related to Visa, sharing a single IP address of ColoCrossing (AS36352).

\subsection{IP Address Sharing}

As explained in the case studies, similar GSDs reuse IP addresses. 
We analyzed the IP addresses used for domain names of the 2,374 clusters for which A records existed in passive DNS out of 2,874 total clusters. 
There are 1,554 (65.5\%) clusters whose domain names share one or two IP addresses. 
On the other hand, a few domain names in some clusters shared a large number of IP addresses. 
For example, four domain names in the same cluster, which targeted Sparkasse, a German financial institution, shared 265 IP addresses. 
There are 426 clusters where IP addresses associated with domain names outnumbered domain names. 
Although some GSDs can be detected by finding domain names associated with IP addresses of known phishing domain names in passive DNS, it is not a comprehensive method because not all GSDs share IP addresses. 
Our proposed system, which analyzes domain name strings alone, can comprehensively detect GSDs, including domain names that share many IPs with a few domain names.

\subsection{Edit Distance}

\begin{figure}[!t]
\centering
\includegraphics[width=\linewidth]
{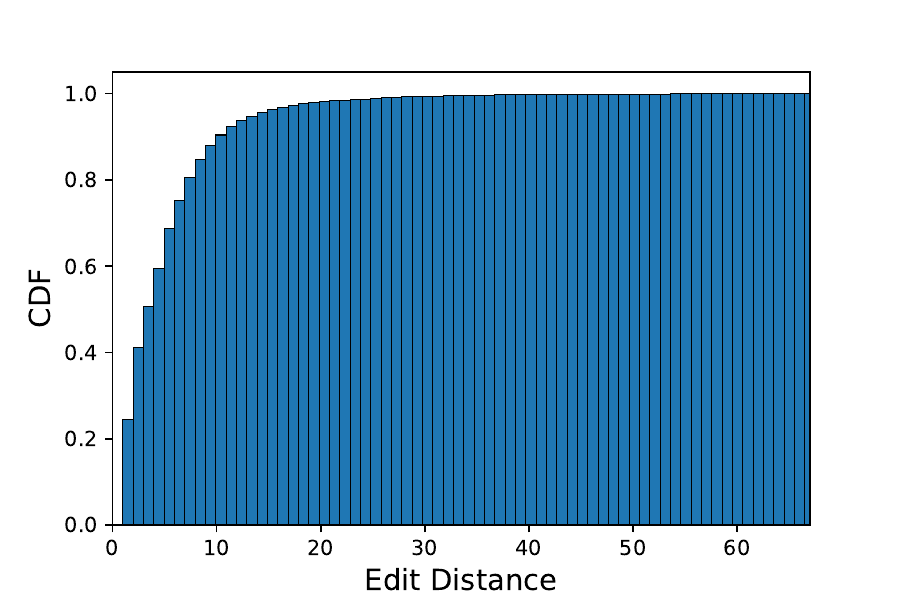}
\caption{Average edit distance between domain names in each cluster.}
\label{fig:editdistance}
\end{figure}

We used the Damerau-Levenshtein distance as the edit distance between domain names for each cluster to analyze the similarity of their appearance.
 We calculated the edit distance for all domain name pairs in each cluster and showed their average in Figure~\ref{fig:editdistance}. 
There are 693 clusters (4,759 domain names) whose average edit distance was less than two. 
Even when trying to find domain names with close edit distance to known phishing domain names, the number of GSDs we can detect is limited. 
Our proposed system can detect GSDs focusing on the similarity of domain names, even if the edit distance increases due to letters and words being replaced, deleted, or added.

\subsection{Phishing Targeted Brands}

\begin{table}[!t]
\centering
\caption{Top 10 categories.}
\label{tab_category}
\begin{tabular}{lr}
\toprule
Category           & \# of domains \\
\midrule
Credit Card        & 117,454 \\
Logistics          & 17,943  \\
Telecommunications & 17,232  \\
Social Networks    & 2,931   \\
Bank               & 2,740   \\
Crypto             & 1,842   \\
Software           & 1,768   \\
E-commerce         & 1,195   \\
Government         & 770    \\
News               & 507    \\
Other              & 1,261  \\
\bottomrule
\end{tabular}
\end{table}

\begin{table}
\centering
\caption{Top 10 countries.}
\label{tab_country}
\begin{tabular}{lr}
\toprule
Country & \# of domains \\
\midrule
Japan      & 149,529 \\
United States      & 12,188  \\
France      & 683    \\
United Kingdom      & 604    \\
Spain      & 409    \\
China      & 321    \\
Turkey      & 305    \\
Italy      & 209    \\
Poland      & 197    \\
Colombia      & 152    \\
Other   & 1,046  \\
\bottomrule
\end{tabular}
\end{table}

\begin{table}[!t]
\centering
\caption{Top 10 brands.}
\label{tab_brand}
\begin{tabular}{llr}
\toprule
Brand              & Country    & \# of domains \\
\midrule
Credit card A      & Japan      & 88,970  \\
Logistics A        & Japan      & 17,345  \\
Telecommunication A & Japan      & 14,905  \\
Credit card B      & Japan      & 14,631  \\
Credit card C      & Japan      & 10,900  \\
Social networks A  & United States      & 2,175   \\
Credit card D      & Japan      & 1,526   \\
Crypto wallet A           & United States      & 1,454   \\
Telecommunication B & United States      & 1,316   \\
E-commerce A       & United States      & 956      \\
\bottomrule
\end{tabular}
\end{table}

We analyzed brands targeted by phishing campaigns.
We extracted brand information associated with 205k domain names from phishing TI and screenshots of crawling results. As a result, we found that 265 brands were imitated in 165,643 domain names, excluding those that could not be identified.
We investigated these brands and confirmed categories and countries in which they primarily provide services. 
The top 10 categories are shown in Table~\ref{tab_category}. 
The Other category includes consumer electronics, gaming, streaming, gambling, and energy. 
The Credit Card category includes 20 brands and accounts for 70.9\% of all domain names. 
Phishing sites targeting financial institutions, telecommunications, and postal services were frequent in many countries. 
Most domain names in the government category were related to tax payments, such as the revenue services and tax agencies in the United States, Australia, Japan, Turkey, France, and the United Kingdom.

Table~\ref{tab_country} lists the top 10 countries (35 countries in total) to which each brand belongs, ranked by the number of domain names. 
In addition, Table~\ref{tab_brand} shows the top 10 brands and countries to which they belong. 
Domain names targeting brands in Japan are the most common, accounting for 90.3\% of the total. 
We investigated brands of all domain names reported on PhishTank and OpenPhish during the same period and found that the United States accounted for 69.7\%, while Japan accounted for 8.4\%. 
Phishing using GSDs has been observed worldwide. 
However, compared with the overall phishing trend, they are particularly prevalent in Japan.

We analyzed 1,594 clusters whose domain names were associated with at least one brand and found that 88\% of the clusters have a single brand. 
The average number of brands per cluster was 1.27.
There were some clusters whose domain names included similar strings to a single brand but targeted multiple brands.
In other words, attackers created GSDs to imitate a brand and used them for phishing for unrelated brands.

\section{Limitations}
In our evaluation of the proposed system, we utilized a verified dataset and conducted appropriate experiments; however, there are certain limitations to consider. 
The system identifies GSDs potentially owned by attackers by examining their similarity to known phishing domain names. 
It is crucial to recognize that not all GSDs prepared by attackers are used for actual phishing sites.
Contrary to previous studies, which required accessing websites to detect phishing content, our proposed system can identify domain names without the need for direct access. 
By solely analyzing domain name strings, our system can efficiently detect domain names prepared by attackers, even before the phishing sites become active.

The proposed system can only detect GSDs that are similar to known phishing domain names, which means that it cannot detect entirely new GSDs, such as those that use new squatting techniques. As described in Section~\ref{sec:duration}, many of the similar GSDs remain active for more than one day. However, if we can observe the early stages of the emergence of phishing sites using new GSDs, we can detect the remaining GSDs that are similar to the domain names.
Furthermore, similar issues occur with existing systems shown in Section~\ref{sec:comparative}. These systems require additional manual effort for brand selection and training when new brands are targeted. In contrast, the proposed system automatically detects the latest phishing-related domain names based on the extracted similarities from Phishing TI. Therefore, our system can respond to the emergence of new brands without the need for frequent additional training.
We conducted experiments using various phishing TIs and URL inspection services as ground truth to validate a diverse set of phishing domain names. However, due to the lack of a comprehensive phishing feed, the potential for false negatives remains. Also, the output of our system may be limited to domain names similar to those listed in the phishing TI.

In this study, we did not perform an analysis of GSDs employing internationalized domain names (IDNs) due to the limited number of observations available for model training. For instance, within a span of 150 days of phishing TI, only approximately 400 IDNs were encountered, and none of these instances involved squatting domain names that led to brand name misidentification. Nevertheless, should GSDs with IDNs arise in the future, our proposed system can detect them by converting Unicode characters to the corresponding ASCII characters (e.g., ã to a) before extracting features. Furthermore, the detection of GSDs incorporating non-English brand names within IDNs can be achieved by fine-tuning a multilingual pre-trained model.

\section{Related Work}

In recent years, domain squatting has emerged as a significant threat to Internet security, with attackers exploiting various techniques to conduct malicious activities. Consequently, numerous studies have been undertaken to investigate and address this issue. For example, Agten et al.\cite{AgtenJPN15} conducted a comprehensive longitudinal study on typosquatting by creating similar strings from legitimate domain names on the basis of edit distance. They found that strict policies and accessible dispute-resolution procedures can effectively reduce typosquatting abuse. 
Similarly, Nikiforakis et al.\cite{NikiforakisAMDPJ13} explored the prevalence of bitsquatting in domain squatting and highlighted its common use in malicious activities such as drive-by download attacks and deception-based software installations.

Furthermore, Simpson et al.\cite{SimpsonMC20} investigated visually impersonating domain names (VIDNs) in business email compromise (BEC) frauds, demonstrating that their implementation of countermeasures has led to a decline in new VIDN registrations by criminals.
Kintis et al.\cite{KintisMLCGPNA17} investigated combosquatting, a technique that combines recognizable brand names with other keywords to create malicious domains. Their findings revealed that combosquatting domains are far more prevalent than typosquatting domains, emphasizing the need for more robust detection mechanisms.
While these studies primarily focus on identifying domain names that are easily mistaken for legitimate ones, our research aims to detect domain names generated by multiple squatting techniques to evade existing detection methods, based on their similarity to known phishing domain names.

Several studies have also targeted the detection of homograph domains, including IDNs, using pairs of Unicode and ASCII characters~\cite{QuinkertLRKH19,Suzuki0YMG19}. Chiba et al.\cite{0001HKSGA19} proposed a system for detecting and scoring deceptive IDNs, incorporating measurement studies and online surveys to evaluate the effectiveness of their scoring metric and suggest practical countermeasures. 
Other research has focused on phishing attacks that target specific industries or events, such as online banking and phishing campaigns ralated to COVID-19 \cite{DamKBS19,KumarGTSL21,XiaNK0Y21}.

Previous studies have also explored the identification of phishing domain names using features specific to certificates in Certificate Transparency (CT) logs and passive DNS traffic. 
Drichel et al.\cite{DrichelDBM21} proposed a detection pipeline capable of performing retrospective analysis and live classification of certificates published in CT logs, using machine learning models to identify phishing domain names. 
Sabah et al.\cite{SabahNBC22} developed an approach to detect phishing by extracting various features from CT logs, certificate-based characteristics, passive DNS traffic, and lexical characteristics. Their live experiments identified phishing domains days before they were detected by Google Safe Browsing and VirusTotal.
Our proposed system aims to provide a comprehensive solution for detecting GSDs by analyzing linguistic characteristics using a language model, without relying on specific input formats. This approach ensures a more robust and versatile detection mechanism in the evolving landscape of domain squatting techniques.

Several systems have been proposed for detecting phishing sites based on their appearance, such as VisualPhishNet~\cite{AbdelnabiKF20}, PhishPedia~\cite{LinLDNCLSZD21}, and PhishIntention~\cite{Liu0YNDD22}. 
These systems use deep learning-based image recognition techniques to compare screenshot images obtained by web crawlers with those of legitimate web pages. 
By leveraging such visual comparison and logo recognition, these systems can accurately detect phishing sites that resemble legitimate ones.
However, the effectiveness of these systems is limited by their dependence on web crawlers.
Specifically, some phishing sites use cloaking techniques that make them inaccessible to web crawlers, making it difficult to identify all phishing sites.
In this study, we used PhishPedia to validate whether the proposed system detected actual phishing sites. 
Consequently, it is essential to implement the proposed system in conjunction with these web crawling systems to effectively mitigate phishing attacks.

\section{Conclusion}
In conclusion, the increasing sophistication of domain squatting techniques requires a more advanced approach to detecting and preventing phishing attacks. In response to this challenge, we have proposed a system called \textsc{PhishReplicant}, which focuses on detecting generated squatting domains (GSDs) that employ a combination of squatting techniques. By leveraging linguistic similarities between known malicious domain names, \textsc{PhishReplicant} can efficiently identify GSDs without the need for frequent model updates.
We have demonstrated the effectiveness of our proposed system by analyzing certificate transparency (CT) logs, lists of newly registered domain names, and passive DNS data observed on 66 DNS cache servers across 18 countries. Over a four-week period, \textsc{PhishReplicant} successfully detected 3,498 GSDs, with 2,821 of these being used for phishing sites within one month of detection. Furthermore, our system outperformed baseline systems in terms of both detection accuracy and the number of identified domain names that did not include exact brand names.
An in-depth analysis of 205k GSDs collected over a 150-day period revealed that these domains targeted 265 brands in 35 countries, with a particular focus on financial institutions and specific geographic regions. This underscores the importance of employing advanced detection system like \textsc{PhishReplicant} to protect users from evolving phishing threats. By continuously monitoring and identifying GSDs, we can proactively prevent phishing attacks before they have the opportunity to cause harm or shortly after they emerge, thereby enhancing overall web security.

\balance
\bibliographystyle{ACM-Reference-Format}

\bibliography{bib}


\begin{thebibliography}{44}


\ifx \showCODEN    \undefined \def \showCODEN     #1{\unskip}     \fi
\ifx \showDOI      \undefined \def \showDOI       #1{#1}\fi
\ifx \showISBNx    \undefined \def \showISBNx     #1{\unskip}     \fi
\ifx \showISBNxiii \undefined \def \showISBNxiii  #1{\unskip}     \fi
\ifx \showISSN     \undefined \def \showISSN      #1{\unskip}     \fi
\ifx \showLCCN     \undefined \def \showLCCN      #1{\unskip}     \fi
\ifx \shownote     \undefined \def \shownote      #1{#1}          \fi
\ifx \showarticletitle \undefined \def \showarticletitle #1{#1}   \fi
\ifx \showURL      \undefined \def \showURL       {\relax}        \fi
\providecommand\bibfield[2]{#2}
\providecommand\bibinfo[2]{#2}
\providecommand\natexlab[1]{#1}
\providecommand\showeprint[2][]{arXiv:#2}

\bibitem[phi(2020)]%
        {phishingcatcher}
 \bibinfo{year}{2020}\natexlab{}.
\newblock \bibinfo{title}{{GitHub - x0rz/phishing\_catcher}}.
\newblock
\newblock
\urldef\tempurl%
\url{https://github.com/x0rz/phishing_catcher}
\showURL{%
\tempurl}


\bibitem[str(2021)]%
        {streamingphish}
 \bibinfo{year}{2021}\natexlab{}.
\newblock \bibinfo{title}{{GitHub - wesleyraptor/streamingphish}}.
\newblock
\newblock
\urldef\tempurl%
\url{https://github.com/wesleyraptor/streamingphish}
\showURL{%
\tempurl}


\bibitem[ctl(2023)]%
        {ctlog}
 \bibinfo{year}{2023}\natexlab{}.
\newblock \bibinfo{title}{{Certificate Transparency}}.
\newblock \bibinfo{howpublished}{\url{https://certificate.transparency.dev/}}.
\newblock


\bibitem[cdp(2023)]%
        {cdp}
 \bibinfo{year}{2023}\natexlab{}.
\newblock \bibinfo{title}{{Chrome DevTools Protocol}}.
\newblock
  \bibinfo{howpublished}{\url{https://chromedevtools.github.io/devtools-protocol/}}.
\newblock


\bibitem[fai(2023)]%
        {faiss}
 \bibinfo{year}{2023}\natexlab{}.
\newblock \bibinfo{title}{{Faiss documentation}}.
\newblock \bibinfo{howpublished}{\url{https://faiss.ai/}}.
\newblock


\bibitem[dns(2023a)]%
        {dnsdb}
 \bibinfo{year}{2023}\natexlab{a}.
\newblock \bibinfo{title}{{Farsight DNSDB}}.
\newblock
  \bibinfo{howpublished}{\url{https://www.farsightsecurity.com/solutions/dnsdb/}}.
\newblock


\bibitem[dns(2023b)]%
        {dnstwist}
 \bibinfo{year}{2023}\natexlab{b}.
\newblock \bibinfo{title}{{GitHub - elceef/dnstwist}}.
\newblock
\newblock
\urldef\tempurl%
\url{https://github.com/elceef/dnstwist}
\showURL{%
\tempurl}


\bibitem[ope(2023)]%
        {openphish}
 \bibinfo{year}{2023}\natexlab{}.
\newblock \bibinfo{title}{{OpenPhish}}.
\newblock
\newblock
\urldef\tempurl%
\url{https://openphish.com/}
\showURL{%
\tempurl}


\bibitem[phi(2023)]%
        {phishtank}
 \bibinfo{year}{2023}\natexlab{}.
\newblock \bibinfo{title}{{PhishTank}}.
\newblock
\newblock
\urldef\tempurl%
\url{https://phishtank.org/}
\showURL{%
\tempurl}


\bibitem[sbe(2023)]%
        {sbert}
 \bibinfo{year}{2023}\natexlab{}.
\newblock \bibinfo{title}{{SentenceTransformers Documentation}}.
\newblock \bibinfo{howpublished}{\url{https://www.sbert.net/}}.
\newblock


\bibitem[url(2023)]%
        {urlscan}
 \bibinfo{year}{2023}\natexlab{}.
\newblock \bibinfo{title}{{urlscan.io}}.
\newblock \bibinfo{howpublished}{\url{https://urlscan.io/}}.
\newblock


\bibitem[vir(2023)]%
        {virustotal}
 \bibinfo{year}{2023}\natexlab{}.
\newblock \bibinfo{title}{{VirusTotal}}.
\newblock \bibinfo{howpublished}{\url{https://www.virustotal.com/}}.
\newblock


\bibitem[zon(2023)]%
        {zonefiles}
 \bibinfo{year}{2023}\natexlab{}.
\newblock \bibinfo{title}{{Zonefiles}}.
\newblock \bibinfo{howpublished}{\url{https://zonefiles.io/}}.
\newblock


\bibitem[Abdelnabi et~al\mbox{.}(2020)]%
        {AbdelnabiKF20}
\bibfield{author}{\bibinfo{person}{Sahar Abdelnabi}, \bibinfo{person}{Katharina
  Krombholz}, {and} \bibinfo{person}{Mario Fritz}.}
  \bibinfo{year}{2020}\natexlab{}.
\newblock \showarticletitle{VisualPhishNet: Zero-Day Phishing Website Detection
  by Visual Similarity}. In \bibinfo{booktitle}{\emph{{CCS} '20: 2020 {ACM}
  {SIGSAC} Conference on Computer and Communications Security, Virtual Event,
  USA, November 9-13, 2020}}, \bibfield{editor}{\bibinfo{person}{Jay Ligatti},
  \bibinfo{person}{Xinming Ou}, \bibinfo{person}{Jonathan Katz}, {and}
  \bibinfo{person}{Giovanni Vigna}} (Eds.). \bibinfo{publisher}{{ACM}},
  \bibinfo{pages}{1681--1698}.
\newblock


\bibitem[Agten et~al\mbox{.}(2015)]%
        {AgtenJPN15}
\bibfield{author}{\bibinfo{person}{Pieter Agten}, \bibinfo{person}{Wouter
  Joosen}, \bibinfo{person}{Frank Piessens}, {and} \bibinfo{person}{Nick
  Nikiforakis}.} \bibinfo{year}{2015}\natexlab{}.
\newblock \showarticletitle{Seven Months' Worth of Mistakes: {A} Longitudinal
  Study of Typosquatting Abuse}. In \bibinfo{booktitle}{\emph{22nd Annual
  Network and Distributed System Security Symposium, {NDSS} 2015, San Diego,
  California, USA, February 8-11, 2015}}. \bibinfo{publisher}{The Internet
  Society}.
\newblock


\bibitem[Azad et~al\mbox{.}(2020)]%
        {AzadSLN20}
\bibfield{author}{\bibinfo{person}{Babak~Amin Azad}, \bibinfo{person}{Oleksii
  Starov}, \bibinfo{person}{Pierre Laperdrix}, {and} \bibinfo{person}{Nick
  Nikiforakis}.} \bibinfo{year}{2020}\natexlab{}.
\newblock \showarticletitle{Web Runner 2049: Evaluating Third-Party Anti-bot
  Services}. In \bibinfo{booktitle}{\emph{Detection of Intrusions and Malware,
  and Vulnerability Assessment - 17th International Conference, {DIMVA} 2020,
  Lisbon, Portugal, June 24-26, 2020, Proceedings}},
  \bibfield{editor}{\bibinfo{person}{Cl{\'{e}}mentine Maurice},
  \bibinfo{person}{Leyla Bilge}, \bibinfo{person}{Gianluca Stringhini}, {and}
  \bibinfo{person}{Nuno Neves}} (Eds.), Vol.~\bibinfo{volume}{12223}.
  \bibinfo{publisher}{Springer}, \bibinfo{pages}{135--159}.
\newblock


\bibitem[Chiba et~al\mbox{.}(2019)]%
        {0001HKSGA19}
\bibfield{author}{\bibinfo{person}{Daiki Chiba}, \bibinfo{person}{Ayako~Akiyama
  Hasegawa}, \bibinfo{person}{Takashi Koide}, \bibinfo{person}{Yuta Sawabe},
  \bibinfo{person}{Shigeki Goto}, {and} \bibinfo{person}{Mitsuaki Akiyama}.}
  \bibinfo{year}{2019}\natexlab{}.
\newblock \showarticletitle{DomainScouter: Understanding the Risks of Deceptive
  IDNs}. In \bibinfo{booktitle}{\emph{22nd International Symposium on Research
  in Attacks, Intrusions and Defenses, {RAID} 2019, Chaoyang District, Beijing,
  China, September 23-25, 2019}}. \bibinfo{publisher}{{USENIX} Association},
  \bibinfo{pages}{413--426}.
\newblock


\bibitem[Dam et~al\mbox{.}(2019)]%
        {DamKBS19}
\bibfield{author}{\bibinfo{person}{Tobias Dam}, \bibinfo{person}{Lukas~Daniel
  Klausner}, \bibinfo{person}{Damjan Buhov}, {and} \bibinfo{person}{Sebastian
  Schrittwieser}.} \bibinfo{year}{2019}\natexlab{}.
\newblock \showarticletitle{Large-Scale Analysis of Pop-Up Scam on
  Typosquatting URLs}. In \bibinfo{booktitle}{\emph{Proceedings of the 14th
  International Conference on Availability, Reliability and Security, {ARES}
  2019, Canterbury, UK, August 26-29, 2019}}. \bibinfo{publisher}{{ACM}},
  \bibinfo{pages}{53:1--53:9}.
\newblock


\bibitem[Devlin et~al\mbox{.}(2019)]%
        {DevlinCLT19}
\bibfield{author}{\bibinfo{person}{Jacob Devlin}, \bibinfo{person}{Ming{-}Wei
  Chang}, \bibinfo{person}{Kenton Lee}, {and} \bibinfo{person}{Kristina
  Toutanova}.} \bibinfo{year}{2019}\natexlab{}.
\newblock \showarticletitle{{BERT:} Pre-training of Deep Bidirectional
  Transformers for Language Understanding}. In
  \bibinfo{booktitle}{\emph{Proceedings of the 2019 Conference of the North
  American Chapter of the Association for Computational Linguistics: Human
  Language Technologies, {NAACL-HLT} 2019, Minneapolis, MN, USA, June 2-7,
  2019, Volume 1 (Long and Short Papers)}},
  \bibfield{editor}{\bibinfo{person}{Jill Burstein}, \bibinfo{person}{Christy
  Doran}, {and} \bibinfo{person}{Thamar Solorio}} (Eds.).
  \bibinfo{publisher}{Association for Computational Linguistics},
  \bibinfo{pages}{4171--4186}.
\newblock


\bibitem[Drichel et~al\mbox{.}(2021)]%
        {DrichelDBM21}
\bibfield{author}{\bibinfo{person}{Arthur Drichel}, \bibinfo{person}{Vincent
  Drury}, \bibinfo{person}{Justus von Brandt}, {and} \bibinfo{person}{Ulrike
  Meyer}.} \bibinfo{year}{2021}\natexlab{}.
\newblock \showarticletitle{Finding Phish in a Haystack: {A} Pipeline for
  Phishing Classification on Certificate Transparency Logs}. In
  \bibinfo{booktitle}{\emph{{ARES} 2021: The 16th International Conference on
  Availability, Reliability and Security, Vienna, Austria, August 17-20,
  2021}}, \bibfield{editor}{\bibinfo{person}{Delphine Reinhardt} {and}
  \bibinfo{person}{Tilo M{\"{u}}ller}} (Eds.). \bibinfo{publisher}{{ACM}},
  \bibinfo{pages}{59:1--59:12}.
\newblock


\bibitem[Kim et~al\mbox{.}(2021)]%
        {KimCKDSAD21}
\bibfield{author}{\bibinfo{person}{Doowon Kim}, \bibinfo{person}{Haehyun Cho},
  \bibinfo{person}{Yonghwi Kwon}, \bibinfo{person}{Adam Doup{\'{e}}},
  \bibinfo{person}{Sooel Son}, \bibinfo{person}{Gail{-}Joon Ahn}, {and}
  \bibinfo{person}{Tudor Dumitras}.} \bibinfo{year}{2021}\natexlab{}.
\newblock \showarticletitle{Security Analysis on Practices of Certificate
  Authorities in the {HTTPS} Phishing Ecosystem}. In
  \bibinfo{booktitle}{\emph{{ASIA} {CCS} '21: {ACM} Asia Conference on Computer
  and Communications Security, Virtual Event, Hong Kong, June 7-11, 2021}},
  \bibfield{editor}{\bibinfo{person}{Jiannong Cao}, \bibinfo{person}{Man~Ho
  Au}, \bibinfo{person}{Zhiqiang Lin}, {and} \bibinfo{person}{Moti Yung}}
  (Eds.). \bibinfo{publisher}{{ACM}}, \bibinfo{pages}{407--420}.
\newblock


\bibitem[Kintis et~al\mbox{.}(2017)]%
        {KintisMLCGPNA17}
\bibfield{author}{\bibinfo{person}{Panagiotis Kintis}, \bibinfo{person}{Najmeh
  Miramirkhani}, \bibinfo{person}{Charles Lever}, \bibinfo{person}{Yizheng
  Chen}, \bibinfo{person}{Rosa~Romero G{\'{o}}mez}, \bibinfo{person}{Nikolaos
  Pitropakis}, \bibinfo{person}{Nick Nikiforakis}, {and} \bibinfo{person}{Manos
  Antonakakis}.} \bibinfo{year}{2017}\natexlab{}.
\newblock \showarticletitle{Hiding in Plain Sight: {A} Longitudinal Study of
  Combosquatting Abuse}. In \bibinfo{booktitle}{\emph{Proceedings of the 2017
  {ACM} {SIGSAC} Conference on Computer and Communications Security, {CCS}
  2017, Dallas, TX, USA, October 30 - November 03, 2017}},
  \bibfield{editor}{\bibinfo{person}{Bhavani Thuraisingham},
  \bibinfo{person}{David Evans}, \bibinfo{person}{Tal Malkin}, {and}
  \bibinfo{person}{Dongyan Xu}} (Eds.). \bibinfo{publisher}{{ACM}},
  \bibinfo{pages}{569--586}.
\newblock


\bibitem[Kumar et~al\mbox{.}(2021)]%
        {KumarGTSL21}
\bibfield{author}{\bibinfo{person}{Neeraj Kumar}, \bibinfo{person}{Sukhada
  Ghewari}, \bibinfo{person}{Harshal Tupsamudre}, \bibinfo{person}{Manish
  Shukla}, {and} \bibinfo{person}{Sachin Lodha}.}
  \bibinfo{year}{2021}\natexlab{}.
\newblock \showarticletitle{When Diversity Meets Hostility: {A} Study of Domain
  Squatting Abuse in Online Banking}. In \bibinfo{booktitle}{\emph{{APWG}
  Symposium on Electronic Crime Research, eCrime 2021, Boston, MA, USA,
  December 1-3, 2021}}. \bibinfo{publisher}{{IEEE}}, \bibinfo{pages}{1--15}.
\newblock


\bibitem[Lin et~al\mbox{.}(2021)]%
        {LinLDNCLSZD21}
\bibfield{author}{\bibinfo{person}{Yun Lin}, \bibinfo{person}{Ruofan Liu},
  \bibinfo{person}{Dinil~Mon Divakaran}, \bibinfo{person}{Jun~Yang Ng},
  \bibinfo{person}{Qing~Zhou Chan}, \bibinfo{person}{Yiwen Lu},
  \bibinfo{person}{Yuxuan Si}, \bibinfo{person}{Fan Zhang}, {and}
  \bibinfo{person}{Jin~Song Dong}.} \bibinfo{year}{2021}\natexlab{}.
\newblock \showarticletitle{Phishpedia: {A} Hybrid Deep Learning Based Approach
  to Visually Identify Phishing Webpages}. In \bibinfo{booktitle}{\emph{30th
  {USENIX} Security Symposium, {USENIX} Security 2021, August 11-13, 2021}},
  \bibfield{editor}{\bibinfo{person}{Michael Bailey} {and}
  \bibinfo{person}{Rachel Greenstadt}} (Eds.). \bibinfo{publisher}{{USENIX}
  Association}, \bibinfo{pages}{3793--3810}.
\newblock


\bibitem[Liu et~al\mbox{.}(2021)]%
        {Liu0LLDS21}
\bibfield{author}{\bibinfo{person}{Mingxuan Liu}, \bibinfo{person}{Yiming
  Zhang}, \bibinfo{person}{Baojun Liu}, \bibinfo{person}{Zhou Li},
  \bibinfo{person}{Haixin Duan}, {and} \bibinfo{person}{Donghong Sun}.}
  \bibinfo{year}{2021}\natexlab{}.
\newblock \showarticletitle{Detecting and Characterizing {SMS} Spearphishing
  Attacks}. In \bibinfo{booktitle}{\emph{{ACSAC} '21: Annual Computer Security
  Applications Conference, Virtual Event, USA, December 6 - 10, 2021}}.
  \bibinfo{publisher}{{ACM}}, \bibinfo{pages}{930--943}.
\newblock


\bibitem[Liu et~al\mbox{.}(2022)]%
        {Liu0YNDD22}
\bibfield{author}{\bibinfo{person}{Ruofan Liu}, \bibinfo{person}{Yun Lin},
  \bibinfo{person}{Xianglin Yang}, \bibinfo{person}{Siang~Hwee Ng},
  \bibinfo{person}{Dinil~Mon Divakaran}, {and} \bibinfo{person}{Jin~Song
  Dong}.} \bibinfo{year}{2022}\natexlab{}.
\newblock \showarticletitle{Inferring Phishing Intention via Webpage Appearance
  and Dynamics: {A} Deep Vision Based Approach}. In
  \bibinfo{booktitle}{\emph{31st {USENIX} Security Symposium, {USENIX} Security
  2022, Boston, MA, USA, August 10-12, 2022}},
  \bibfield{editor}{\bibinfo{person}{Kevin R.~B. Butler} {and}
  \bibinfo{person}{Kurt Thomas}} (Eds.). \bibinfo{publisher}{{USENIX}
  Association}, \bibinfo{pages}{1633--1650}.
\newblock


\bibitem[Loyola et~al\mbox{.}(2020)]%
        {LoyolaGKWS20}
\bibfield{author}{\bibinfo{person}{Pablo Loyola}, \bibinfo{person}{Kugamoorthy
  Gajananan}, \bibinfo{person}{Hirokuni Kitahara}, \bibinfo{person}{Yuji
  Watanabe}, {and} \bibinfo{person}{Fumiko Satoh}.}
  \bibinfo{year}{2020}\natexlab{}.
\newblock \showarticletitle{Automating Domain Squatting Detection Using
  Representation Learning}. In \bibinfo{booktitle}{\emph{2020 {IEEE}
  International Conference on Big Data {(IEEE} BigData 2020), Atlanta, GA, USA,
  December 10-13, 2020}}, \bibfield{editor}{\bibinfo{person}{Xintao Wu},
  \bibinfo{person}{Chris Jermaine}, \bibinfo{person}{Li~Xiong},
  \bibinfo{person}{Xiaohua Hu}, \bibinfo{person}{Olivera Kotevska},
  \bibinfo{person}{Siyuan Lu}, \bibinfo{person}{Weija Xu},
  \bibinfo{person}{Srinivas Aluru}, \bibinfo{person}{Chengxiang Zhai},
  \bibinfo{person}{Eyhab Al{-}Masri}, \bibinfo{person}{Zhiyuan Chen}, {and}
  \bibinfo{person}{Jeff Saltz}} (Eds.). \bibinfo{publisher}{{IEEE}},
  \bibinfo{pages}{1021--1030}.
\newblock


\bibitem[Nakano et~al\mbox{.}(2023)]%
        {nakano2023canary}
\bibfield{author}{\bibinfo{person}{Hiroki Nakano}, \bibinfo{person}{Daiki
  Chiba}, \bibinfo{person}{Takashi Koide}, \bibinfo{person}{Naoki Fukushi},
  \bibinfo{person}{Takeshi Yagi}, \bibinfo{person}{Takeo Hariu},
  \bibinfo{person}{Katsunari Yoshioka}, {and} \bibinfo{person}{Tsutomu
  Matsumoto}.} \bibinfo{year}{2023}\natexlab{}.
\newblock \showarticletitle{Canary in Twitter Mine: Collecting Phishing Reports
  from Experts and Non-experts}. In \bibinfo{booktitle}{\emph{Proceedings of
  the 18th International Conference on Availability, Reliability and Security,
  {ARES} 2023, Benevento, Italy, 29 August 2023- 1 September 2023}}.
  \bibinfo{publisher}{{ACM}}, \bibinfo{pages}{6:1--6:12}.
\newblock
\urldef\tempurl%
\url{https://doi.org/10.1145/3600160.3600163}
\showDOI{\tempurl}


\bibitem[Nikiforakis et~al\mbox{.}(2013)]%
        {NikiforakisAMDPJ13}
\bibfield{author}{\bibinfo{person}{Nick Nikiforakis}, \bibinfo{person}{Steven
  {Van Acker}}, \bibinfo{person}{Wannes Meert}, \bibinfo{person}{Lieven
  Desmet}, \bibinfo{person}{Frank Piessens}, {and} \bibinfo{person}{Wouter
  Joosen}.} \bibinfo{year}{2013}\natexlab{}.
\newblock \showarticletitle{Bitsquatting: exploiting bit-flips for fun, or
  profit?}. In \bibinfo{booktitle}{\emph{22nd International World Wide Web
  Conference, {WWW} '13, Rio de Janeiro, Brazil, May 13-17, 2013}},
  \bibfield{editor}{\bibinfo{person}{Daniel Schwabe},
  \bibinfo{person}{Virg{\'{\i}}lio A.~F. Almeida}, \bibinfo{person}{Hartmut
  Glaser}, \bibinfo{person}{Ricardo Baeza{-}Yates}, {and}
  \bibinfo{person}{Sue~B. Moon}} (Eds.). \bibinfo{pages}{989--998}.
\newblock


\bibitem[Oest et~al\mbox{.}(2019)]%
        {OestSDAWT19}
\bibfield{author}{\bibinfo{person}{Adam Oest}, \bibinfo{person}{Yeganeh
  Safaei}, \bibinfo{person}{Adam Doup{\'{e}}}, \bibinfo{person}{Gail{-}Joon
  Ahn}, \bibinfo{person}{Brad Wardman}, {and} \bibinfo{person}{Kevin Tyers}.}
  \bibinfo{year}{2019}\natexlab{}.
\newblock \showarticletitle{PhishFarm: {A} Scalable Framework for Measuring the
  Effectiveness of Evasion Techniques against Browser Phishing Blacklists}. In
  \bibinfo{booktitle}{\emph{2019 {IEEE} Symposium on Security and Privacy, {SP}
  2019, San Francisco, CA, USA, May 19-23, 2019}}. \bibinfo{publisher}{{IEEE}},
  \bibinfo{pages}{1344--1361}.
\newblock


\bibitem[Oest et~al\mbox{.}(2018)]%
        {OestSDAWW18}
\bibfield{author}{\bibinfo{person}{Adam Oest}, \bibinfo{person}{Yeganeh
  Safaei}, \bibinfo{person}{Adam Doup{\'{e}}}, \bibinfo{person}{Gail{-}Joon
  Ahn}, \bibinfo{person}{Brad Wardman}, {and} \bibinfo{person}{Gary Warner}.}
  \bibinfo{year}{2018}\natexlab{}.
\newblock \showarticletitle{Inside a phisher's mind: Understanding the
  anti-phishing ecosystem through phishing kit analysis}. In
  \bibinfo{booktitle}{\emph{2018 {APWG} Symposium on Electronic Crime Research,
  eCrime 2018, San Diego, CA, USA, May 15-17, 2018}}.
  \bibinfo{publisher}{{IEEE}}, \bibinfo{pages}{1--12}.
\newblock


\bibitem[Peng et~al\mbox{.}(2019)]%
        {PengXQ0VW19}
\bibfield{author}{\bibinfo{person}{Peng Peng}, \bibinfo{person}{Chao Xu},
  \bibinfo{person}{Luke Quinn}, \bibinfo{person}{Hang Hu},
  \bibinfo{person}{Bimal Viswanath}, {and} \bibinfo{person}{Gang Wang}.}
  \bibinfo{year}{2019}\natexlab{}.
\newblock \showarticletitle{What Happens After You Leak Your Password:
  Understanding Credential Sharing on Phishing Sites}. In
  \bibinfo{booktitle}{\emph{Proceedings of the 2019 {ACM} Asia Conference on
  Computer and Communications Security, AsiaCCS 2019, Auckland, New Zealand,
  July 09-12, 2019}}, \bibfield{editor}{\bibinfo{person}{Steven~D. Galbraith},
  \bibinfo{person}{Giovanni Russello}, \bibinfo{person}{Willy Susilo},
  \bibinfo{person}{Dieter Gollmann}, \bibinfo{person}{Engin Kirda}, {and}
  \bibinfo{person}{Zhenkai Liang}} (Eds.). \bibinfo{publisher}{{ACM}},
  \bibinfo{pages}{181--192}.
\newblock


\bibitem[Quinkert et~al\mbox{.}(2019)]%
        {QuinkertLRKH19}
\bibfield{author}{\bibinfo{person}{Florian Quinkert}, \bibinfo{person}{Tobias
  Lauinger}, \bibinfo{person}{William~K. Robertson}, \bibinfo{person}{Engin
  Kirda}, {and} \bibinfo{person}{Thorsten Holz}.}
  \bibinfo{year}{2019}\natexlab{}.
\newblock \showarticletitle{It's Not what It Looks Like: Measuring Attacks and
  Defensive Registrations of Homograph Domains}. In
  \bibinfo{booktitle}{\emph{7th {IEEE} Conference on Communications and Network
  Security, {CNS} 2019, Washington, DC, USA, June 10-12, 2019}}.
  \bibinfo{publisher}{{IEEE}}, \bibinfo{pages}{259--267}.
\newblock


\bibitem[Reimers and Gurevych(2019)]%
        {ReimersG19}
\bibfield{author}{\bibinfo{person}{Nils Reimers} {and} \bibinfo{person}{Iryna
  Gurevych}.} \bibinfo{year}{2019}\natexlab{}.
\newblock \showarticletitle{Sentence-BERT: Sentence Embeddings using Siamese
  BERT-Networks}. In \bibinfo{booktitle}{\emph{Proceedings of the 2019
  Conference on Empirical Methods in Natural Language Processing and the 9th
  International Joint Conference on Natural Language Processing, {EMNLP-IJCNLP}
  2019, Hong Kong, China, November 3-7, 2019}},
  \bibfield{editor}{\bibinfo{person}{Kentaro Inui}, \bibinfo{person}{Jing
  Jiang}, \bibinfo{person}{Vincent Ng}, {and} \bibinfo{person}{Xiaojun Wan}}
  (Eds.). \bibinfo{publisher}{Association for Computational Linguistics},
  \bibinfo{pages}{3980--3990}.
\newblock


\bibitem[Sabah et~al\mbox{.}(2022)]%
        {SabahNBC22}
\bibfield{author}{\bibinfo{person}{Mashael~Al Sabah}, \bibinfo{person}{Mohamed
  Nabeel}, \bibinfo{person}{Yazan Boshmaf}, {and} \bibinfo{person}{Euijin
  Choo}.} \bibinfo{year}{2022}\natexlab{}.
\newblock \showarticletitle{Content-Agnostic Detection of Phishing Domains
  using Certificate Transparency and Passive {DNS}}. In
  \bibinfo{booktitle}{\emph{25th International Symposium on Research in
  Attacks, Intrusions and Defenses, {RAID} 2022, Limassol, Cyprus, October
  26-28, 2022}}. \bibinfo{publisher}{{ACM}}, \bibinfo{pages}{446--459}.
\newblock


\bibitem[Simpson et~al\mbox{.}(2020)]%
        {SimpsonMC20}
\bibfield{author}{\bibinfo{person}{Geoffrey Simpson}, \bibinfo{person}{Tyler
  Moore}, {and} \bibinfo{person}{Richard Clayton}.}
  \bibinfo{year}{2020}\natexlab{}.
\newblock \showarticletitle{Ten years of attacks on companies using visual
  impersonation of domain names}. In \bibinfo{booktitle}{\emph{{APWG} Symposium
  on Electronic Crime Research, eCrime 2020, Boston, MA, USA, November 16-19,
  2020}}. \bibinfo{publisher}{{IEEE}}, \bibinfo{pages}{1--12}.
\newblock


\bibitem[Subramani et~al\mbox{.}(2022)]%
        {SubramaniMSVP22}
\bibfield{author}{\bibinfo{person}{Karthika Subramani},
  \bibinfo{person}{William Melicher}, \bibinfo{person}{Oleksii Starov},
  \bibinfo{person}{Phani Vadrevu}, {and} \bibinfo{person}{Roberto Perdisci}.}
  \bibinfo{year}{2022}\natexlab{}.
\newblock \showarticletitle{PhishInPatterns: measuring elicited user
  interactions at scale on phishing websites}. In
  \bibinfo{booktitle}{\emph{Proceedings of the 22nd {ACM} Internet Measurement
  Conference, {IMC} 2022, Nice, France, October 25-27, 2022}},
  \bibfield{editor}{\bibinfo{person}{Chadi Barakat}, \bibinfo{person}{Cristel
  Pelsser}, \bibinfo{person}{Theophilus~A. Benson}, {and}
  \bibinfo{person}{David Choffnes}} (Eds.). \bibinfo{publisher}{{ACM}},
  \bibinfo{pages}{589--604}.
\newblock


\bibitem[Suzuki et~al\mbox{.}(2019)]%
        {Suzuki0YMG19}
\bibfield{author}{\bibinfo{person}{Hiroaki Suzuki}, \bibinfo{person}{Daiki
  Chiba}, \bibinfo{person}{Yoshiro Yoneya}, \bibinfo{person}{Tatsuya Mori},
  {and} \bibinfo{person}{Shigeki Goto}.} \bibinfo{year}{2019}\natexlab{}.
\newblock \showarticletitle{ShamFinder: An Automated Framework for Detecting
  {IDN} Homographs}. In \bibinfo{booktitle}{\emph{Proceedings of the Internet
  Measurement Conference, {IMC} 2019, Amsterdam, The Netherlands, October
  21-23, 2019}}. \bibinfo{publisher}{{ACM}}, \bibinfo{pages}{449--462}.
\newblock


\bibitem[Szurdi et~al\mbox{.}(2021)]%
        {SzurdiLKNC21}
\bibfield{author}{\bibinfo{person}{Janos Szurdi}, \bibinfo{person}{Meng Luo},
  \bibinfo{person}{Brian Kondracki}, \bibinfo{person}{Nick Nikiforakis}, {and}
  \bibinfo{person}{Nicolas Christin}.} \bibinfo{year}{2021}\natexlab{}.
\newblock \showarticletitle{Where are you taking me?Understanding Abusive
  Traffic Distribution Systems}. In \bibinfo{booktitle}{\emph{{WWW} '21: The
  Web Conference 2021, Virtual Event / Ljubljana, Slovenia, April 19-23,
  2021}}, \bibfield{editor}{\bibinfo{person}{Jure Leskovec},
  \bibinfo{person}{Marko Grobelnik}, \bibinfo{person}{Marc Najork},
  \bibinfo{person}{Jie Tang}, {and} \bibinfo{person}{Leila Zia}} (Eds.).
  \bibinfo{publisher}{{ACM} / {IW3C2}}, \bibinfo{pages}{3613--3624}.
\newblock


\bibitem[Tang et~al\mbox{.}(2022)]%
        {TangML0022}
\bibfield{author}{\bibinfo{person}{Siyuan Tang}, \bibinfo{person}{Xianghang
  Mi}, \bibinfo{person}{Ying Li}, \bibinfo{person}{XiaoFeng Wang}, {and}
  \bibinfo{person}{Kai Chen}.} \bibinfo{year}{2022}\natexlab{}.
\newblock \showarticletitle{Clues in Tweets: Twitter-Guided Discovery and
  Analysis of {SMS} Spam}. In \bibinfo{booktitle}{\emph{Proceedings of the 2022
  {ACM} {SIGSAC} Conference on Computer and Communications Security, {CCS}
  2022, Los Angeles, CA, USA, November 7-11, 2022}},
  \bibfield{editor}{\bibinfo{person}{Heng Yin}, \bibinfo{person}{Angelos
  Stavrou}, \bibinfo{person}{Cas Cremers}, {and} \bibinfo{person}{Elaine Shi}}
  (Eds.). \bibinfo{publisher}{{ACM}}, \bibinfo{pages}{2751--2764}.
\newblock


\bibitem[Wang et~al\mbox{.}(2006)]%
        {WangBWVD06}
\bibfield{author}{\bibinfo{person}{Yi{-}Min Wang}, \bibinfo{person}{Doug Beck},
  \bibinfo{person}{Jeffrey Wang}, \bibinfo{person}{Chad Verbowski}, {and}
  \bibinfo{person}{Brad Daniels}.} \bibinfo{year}{2006}\natexlab{}.
\newblock \showarticletitle{Strider Typo-Patrol: Discovery and Analysis of
  Systematic Typo-Squatting}. In \bibinfo{booktitle}{\emph{2nd Workshop on
  Steps to Reducing Unwanted Traffic on the Internet, SRUTI'06, San Jose, CA,
  USA, July 7, 2006}}, \bibfield{editor}{\bibinfo{person}{Steven~M. Bellovin}}
  (Ed.). \bibinfo{publisher}{{USENIX} Association}.
\newblock


\bibitem[Xia et~al\mbox{.}(2021)]%
        {XiaNK0Y21}
\bibfield{author}{\bibinfo{person}{Pengcheng Xia}, \bibinfo{person}{Mohamed
  Nabeel}, \bibinfo{person}{Issa Khalil}, \bibinfo{person}{Haoyu Wang}, {and}
  \bibinfo{person}{Ting Yu}.} \bibinfo{year}{2021}\natexlab{}.
\newblock \showarticletitle{Identifying and Characterizing {COVID-19} Themed
  Malicious Domain Campaigns}. In \bibinfo{booktitle}{\emph{{CODASPY} '21:
  Eleventh {ACM} Conference on Data and Application Security and Privacy,
  Virtual Event, USA, April 26-28, 2021}},
  \bibfield{editor}{\bibinfo{person}{Anupam Joshi}, \bibinfo{person}{Barbara
  Carminati}, {and} \bibinfo{person}{Rakesh~M. Verma}} (Eds.).
  \bibinfo{publisher}{{ACM}}, \bibinfo{pages}{209--220}.
\newblock


\bibitem[Zeng et~al\mbox{.}(2021)]%
        {ZengCZT21}
\bibfield{author}{\bibinfo{person}{Yuwei Zeng}, \bibinfo{person}{Xunxun Chen},
  \bibinfo{person}{Tianning Zang}, {and} \bibinfo{person}{Haiwei Tsang}.}
  \bibinfo{year}{2021}\natexlab{}.
\newblock \showarticletitle{Winding Path: Characterizing the Malicious
  Redirection in Squatting Domain Names}. In \bibinfo{booktitle}{\emph{Passive
  and Active Measurement - 22nd International Conference, {PAM} 2021, Virtual
  Event, March 29 - April 1, 2021, Proceedings}}
  \emph{(\bibinfo{series}{Lecture Notes in Computer Science},
  Vol.~\bibinfo{volume}{12671})}, \bibfield{editor}{\bibinfo{person}{Oliver
  Hohlfeld}, \bibinfo{person}{Andra Lutu}, {and} \bibinfo{person}{Dave Levin}}
  (Eds.). \bibinfo{publisher}{Springer}, \bibinfo{pages}{93--107}.
\newblock


\bibitem[Zhang et~al\mbox{.}(2021)]%
        {crawlphish}
\bibfield{author}{\bibinfo{person}{Penghui Zhang}, \bibinfo{person}{Adam Oest},
  \bibinfo{person}{Haehyun Cho}, \bibinfo{person}{Zhibo Sun},
  \bibinfo{person}{RC Johnson}, \bibinfo{person}{Brad Wardman},
  \bibinfo{person}{Shaown Sarker}, \bibinfo{person}{Alexandros Kapravelos},
  \bibinfo{person}{Tiffany Bao}, \bibinfo{person}{Ruoyu Wang},
  \bibinfo{person}{Yan Shoshitaishvili}, \bibinfo{person}{Adam Doup{\'{e}}},
  {and} \bibinfo{person}{Gail{-}Joon Ahn}.} \bibinfo{year}{2021}\natexlab{}.
\newblock \showarticletitle{CrawlPhish: Large-scale Analysis of Client-side
  Cloaking Techniques in Phishing}. In \bibinfo{booktitle}{\emph{42nd {IEEE}
  Symposium on Security and Privacy, {SP} 2021, San Francisco, CA, USA, 24-27
  May 2021}}. \bibinfo{publisher}{{IEEE}}, \bibinfo{pages}{1109--1124}.
\newblock


\end{thebibliography}

\cleardoublepage\appendix \onecolumn

\appendix 

\section{Appendix}
\label{sec:appendix}

\begin{figure*}[!b]
\centering
\includegraphics[width=0.8\linewidth,bb=0 0 1070 1251]
{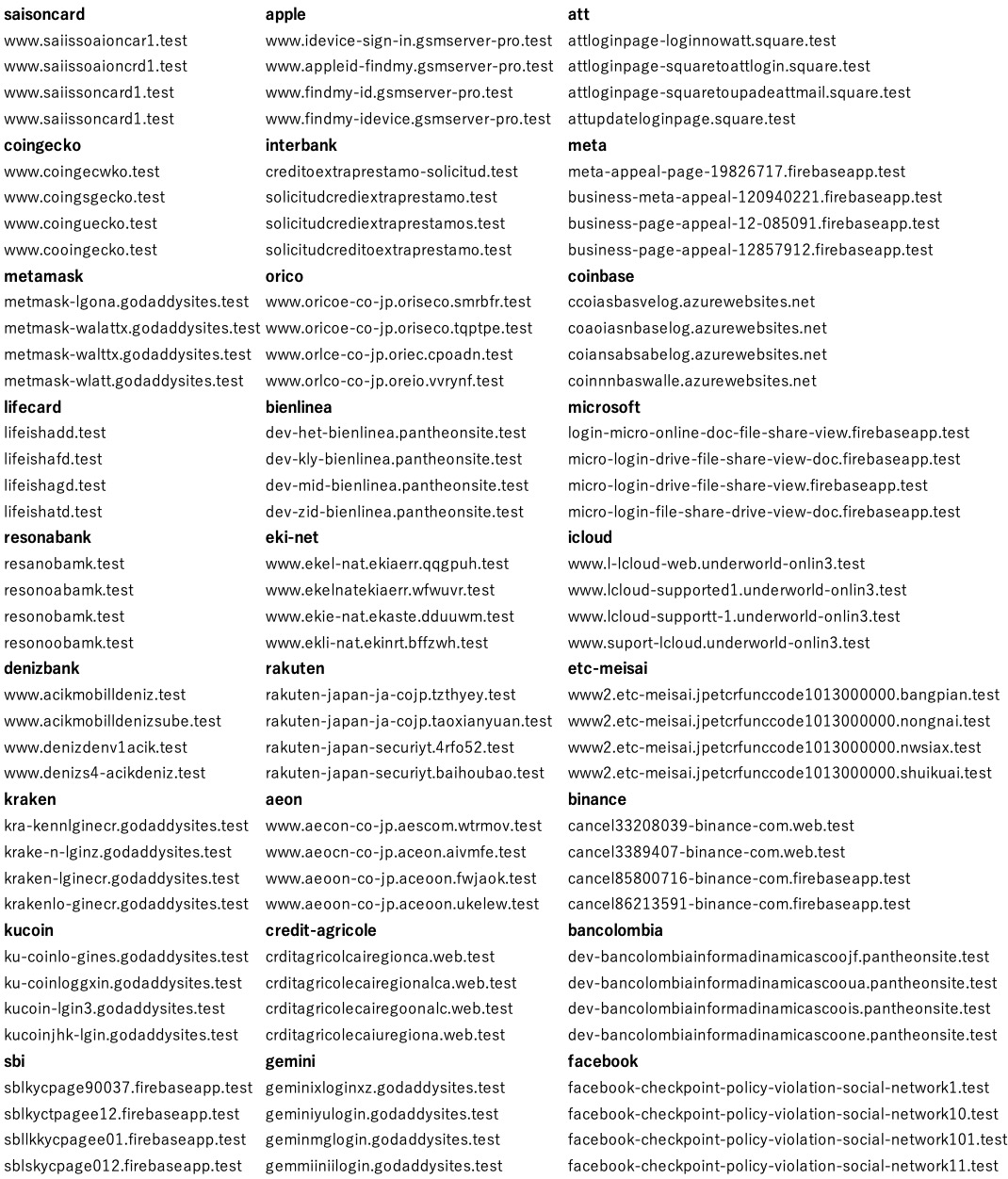}
\caption{Examples of the positives in our dataset used in Section~\ref{sec:eval1}. The TLD of each domain name is converted to \urlstyle{tt}{\url{.test}}.}
\label{fig:dataset}
\end{figure*}

\subsection{An Example of the Dataset}

Figure~\ref{fig:dataset} shows examples of domain names we labeled positives in the dataset, which we used in Section~\ref{sec:eval1}. We extracted four domain names each from sets of similar domain names and investigated their targeted brands.

\end{document}